\documentclass[fleqn,usenatbib]{mnras}

\usepackage{newtxtext,newtxmath}
\usepackage[T1]{fontenc}
\usepackage{graphicx}	
\usepackage{float}
\usepackage{amsmath}	
\usepackage{comment}
\usepackage{multirow}

\DeclareRobustCommand{\VAN}[3]{#2}
\let\VANthebibliography\thebibliography
\def\thebibliography{\DeclareRobustCommand{\VAN}[3]{##3}\VANthebibliography}

\title[OCS emission and source luminosity]{FAUST XXIX. OCS line emission: a new method for measuring the luminosity of embedded protostars in binary systems}

\author[Saury et al.]{
    Guillaume Saury,$^{1}$
    Vittorio Bariosco,$^{2,3}$
    Cecilia Ceccarelli,$^{1}$\thanks{E-mail: cecilia.ceccarelli@univ-grenoble-alpes.fr} 
    Ana L\'opez-Sepulcre,$^{1,4}$
    Layal Chahine,$^{1}$
    \and 
    Marta De Simone,$^5$
    Albert Rimola,$^3$
    Piero Ugliengo,$^2$
    Claire J. Chandler,$^7$
    Nami Sakai,$^8$
    Claudio Codella,$^6$
    \and
    Eleonora Bianchi,$^{6}$
    Lise Boitard--Crépeau,$^1$
    Mathilde Bouvier, ${^9}$
    Romane Le Gal,$^{1,4}$
    Laurent Loinard,$^{10,11,12}$
    \and
    Yoko Oya,$^{13}$
    Linda Podio,$^6$
    Giovanni Sabatini,$^6$
    Charlotte Vastel,$^{14}$
    Ziwei E. Zhang,$^8$
    Satoshi Yamamoto$^{15}$
\\
$^1$Université Grenoble Alpes, CNRS, IPAG, 38000 Grenoble, France\\
$^2$Dipartimento di Chimica, Università degli Studi di Torino, via P. Giuria 7, I-10125 Torino, Italy\\
$^3$Departament de Química, Universitat Autònoma de Barcelona, E-08193 Bellaterra, Catalonia, Spain\\
$^4$Institut de Radioastronomie Millimétrique, 300 rue de la Piscine, Domaine Universitaire, 38406 Saint-Martin d’Hères, France \\
$^5$European Southern Observatory, Karl-Schwarzschild-Strasse 2 D-85748 Garching bei Munchen, Germany\\
$^6$INAF, Osservatorio Astrofisico di Arcetri, Largo E. Fermi 5, 50125 Firenze, Italy\\
$^7$National Radio Astronomy Observatory, PO Box O, Socorro, NM 87801, USA\\
$^8$RIKEN Cluster for Pioneering Research, 2-1, Hirosawa, Wako-shi, Saitama 351-0198, Japan\\
$^9$Leiden Observatory, Leiden University, P.O. Box 9513, 2300 RA Leiden, The Netherlands \\
$^{10}$Instituto de Radioastronomía y Astrofísica, Universidad Nacional Autónoma de México, 58341 Morelia, Mexico\\
$^{11}$Black Hole Initiative at Harvard University, 20 Garden Street, Cambridge, MA 02138, USA\\
$^{12}$David Rockefeller Center for Latin American Studies, Harvard University, 1730 Cambridge Street, Cambridge, MA 02138, USA\\
$^{13}$Yukawa Institute for Theoretical Physics, Kyoto University Oiwake-cho, Kitashirakawa, Sakyo-ku, Kyoto-shi, Kyoto-fu 606-8502, Japan\\
$^{14}$IRAP, Université de Toulouse, CNRS, UPS, CNES, Toulouse, 31400, France\\
$^{15}$SOKENDAI, Shonan Village, Hayama, Kanagawa 240-0193, Japan
}

\date{Accepted 2025 November 26. Received 2025 November 24; in original form 2025 October 22}

\pubyear{2025}

\begin{document}
\label{firstpage}
\pagerange{\pageref{firstpage}--\pageref{lastpage}}
\maketitle

\begin{abstract}
The luminosity of embedded protostars is commonly measured via observations of the dust continuum spectral energy distribution from millimetre to infrared wavelengths.
However, this method cannot be applied to embedded protostars in binary or multiple systems, where their components are usually unresolved over this extended wavelength range.
We propose a new method, based on the idea that a molecule formed (mainly) on the grain surfaces only emits lines in the region where it thermally sublimates from the grain mantles, heated by the photons emitted by the embedded source.
In this respect, carbonyl sulfide (OCS) is an optimal molecule, because of its low binding energy and rotational lines in the millimetre.
We apply the method to the protobinary system NGC1333 IRAS4A, using ALMA high-spatial resolution ($\sim$50 au) observations of the OCS(19-18) line as part of the ALMA Large Program FAUST.
We also present new quantum mechanics calculations of the OCS binding energy distribution, essential for the application of the method.
We found that the two binary components, A1 and A2, have a comparable luminosity within the error bars, 7.5$\pm$2.5 and 7$\pm$1 L$_\odot$, respectively.
We discuss the reliability of the estimated luminosities and the potential of this new method for measuring the luminosity of embedded protostars in binary and multiple systems.
\end{abstract}



\section{Introduction} \label{sec:intro}

The luminosity of a protostar is a fundamental parameter, which is at the base of any theory of star formation \citep[e.g.,][]{Myers1998-Lbol, Wuchterl2001-simulations, Froebrich2006-models, Enoch2009-protostarLum, Offner2011ProtostarsLum}.
It has an impact on countless issues, such as determining the effective accretion rate, likely final star mass, evolutionary status and age, initial luminosity and mass function, episodic accretion, interpretation of a series of dust and gas observations and, more generally, theories of star formation. 
Given the involved temperatures (tens to hundreds K), the luminosity of embedded protostars is obtained by integrating the dust continuum spectral energy distribution (SED) from millimetre to infrared wavelengths.
While this is relatively easy for isolated protostars, thanks to space and ground telescopes in the FIR/IR (e.g. IRAS, ISO, Spitzer, AKARI, Herschel and JWST), where the SED peaks \citep{Evans2009-c2d}, 
measuring the luminosity of each component of embedded protostars in binary or multiple systems is a challenge that has not yet been overcome.
The reason comes from the difficulty of having spatially resolved observations of the system components from the submillimeter to infrared wavelengths, even with the power of the telescopes available nowadays, such as JWST \citep{Gardner2006} and ALMA \citep{Wootten2009}. 

We present here a new method, based on millimetre high-spatial resolution observations of the carbonyl sulphide (OCS) line emission able to disentangle the two binary components, coupled with new quantum mechanics (QM) calculations of the OCS binding energy (BE) distribution, which is the energy needed by an adsorbed molecule to leave the adsorbate surface.  
The basic idea relies on the fact that OCS is an abundant molecule in interstellar ices, actually the most abundant detected S-bearing frozen species \citep[e.g.,][]{Boogert2022, McClure2023-JWSTices, Potapov2025-ices}, and its gaseous abundance in the protostellar envelope is dominated by its thermal (or non-thermal) desorption from the dust grain mantles, because it is inefficiently formed in the gas phase \citep{Loison2012-OCS}.
The region where OCS is sublimated from the grains, therefore, is governed by the protostellar envelope dust temperature profile, which, in turn, depends on the luminosity of the embedded protostar.
Thus, measuring the profile of the OCS line emission around each of the protostellar system multiple components provides an indirect but robust measure of the luminosity of each of them, if the OCS binding energy is well characterised.
In principle, the method applies to whatever molecule is only/mainly formed on the grain surfaces, such as methanol.
However, OCS is optimal because of its relatively low BE, which makes the zone where it is sublimated relatively large and, therefore, easier to map with respect to others molecules, including methanol.

In previous studies, OCS line emission has been used to probe the innermost regions of the infalling rotating envelope.
For example, using the OCS position-velocity (PV) diagram, the presence of an infalling-rotating envelope around IRAS 16293–2422 was suggested \citep{Oya2016}, while only the inner parts at higher velocities were seen around CB68 \citep{Imai2022}. 
Likewise, OCS was observed towards the innermost regions of the disk of L1527, unaffected by shocks \citep{Zhang2024}. 

In the cold prestellar phases of protostar formation, dust grains become covered by dirty ice mantles, whose most abundant component is water in its amorphous form \citep[e.g.,][]{Leger1979, Gibb2004, Boogert2015}.
Recent studies have shown that molecules are adsorbed to these icy surfaces not with a single BE but rather a distribution of BEs, given by the different sites of the icy surface and orientation of the adsorbed molecule \citep[e.g.,][]{Tinacci2022-ammonia, Tinacci2023, bariosco_h2s, Bariosco2025, Groyne2025-BEDs}.
Therefore, to apply the method described above and measure the protostellar luminosity based on the OCS line emission, the OCS BE distribution needs to be known.
However, so far, only single values have been published regarding the OCS BE, which vary from 1570 to 2900 K \citep{McElroy2013, Wakelam2017BE, Penteado2017, Das2018, Perrero2022-SBE}.
In order to fill this gap, in the present work we carried out new QM calculations of the OCS BE distribution.

Armed with the newly computed OCS BE distribution, we applied the method of the OCS line emission to measure the luminosity of the two components of the protobinary system NGC1333 IRAS4A.
To this end, we obtained high-spatial resolution ($\sim 50$ au) of the OCS line emission, as part of the ALMA Large Program "Fifty AU STudy of the chemistry in the disc/envelope system of solar-like protostars" \citep[FAUST;][]{Codella2021}.

The article is organised as follows.
The source background is given in Sec. \ref{sec:source}, the QM calculations of the OCS BE distribution in Sec. \ref{sec:OCS_BE-computations} and the description of the observations in Sec. \ref{sec:observations-results}.
Section \ref{sec:model} reports the model adopted to compute the theoretical OCS line emission and 
Sec. \ref{sec:discussion} presents the  luminosity derived for the two protostars and discusses the reliability of the method. 
Section \ref{conclusions} concludes the article.

\section{The IRAS4A protobinary system} \label{sec:source}

NGC 1333 IRAS4A is part of the IRAS4 system, located in the Perseus Molecular Cloud at a distance of 299$\pm$15 pc  \citep{Zucker2018}, and composed of four objects: 4A, 4B, 4B' and 4C \citep{Sandell1991, Rodriguez1999, Looney2000, Smith2000, Podio2021-jets}.
IRAS4A is a binary system comprising two Class 0 protostars \citep{Lay1995}, IRAS4A1 (hereafter A1) and IRAS4A2 (hereafter A2), separated by about 1.8" \citep[i.e. $\sim$540 au at the source distance;][]{Lopez2017-iras4}. The systemic velocity of the IRAS4 envelope is 7 km/s \citep{Yildiz2012}, while the velocity A1 and A2 are 6.5$\pm$0.3 and 6.9$\pm$0.1 km/s, respectively \citep{Desimone2020-VLA}.
The age of the system is likely between $\sim10^4$ and $\sim10^5$ yr \citep{Maret2002, Kristensen2018-ProtostarsLifetime}.

The total bolometric luminosity of IRAS4A (i.e., A1 + A2) was estimated to be 9 L$_{\odot}$ by \cite{Karska2013-YSOsSED} — see their Appendix C —, using a data set obtained by ground- and space-borne telescopes to construct the SED from 25 $\mu$m to 1.3 mm.
These authors do not report an error on the luminosity estimate, obtained adopting a distance of 235 pc \citep{Hirota2008}, but we assign a conservative 10\% to it.
Once corrected for the new measurement of distance \citep[299$\pm$15 pc:][]{Zucker2018}, the total bolometric luminosity becomes 14.5$\pm$1.5 L$_{\odot}$\footnote{The same luminosity is reported by \cite{vanDishoeck2025} for A1 and A2 separately, probably a typing error.}.
It is important to emphasize that the bolometric luminosity of each binary component is unknown because of the lack of observations disentangling them at infrared wavelengths.

The dust continuum emission in the millimetre was found to be optically thick \citep{Maury2019, Desimone2020-VLA} so that the molecular line emission from the innermost parts is completely absorbed by the dust around A1 and partially around A2 \citep{Desimone2020-VLA, Desimone2022-ices}. 
Despite the dust being optically thick, IRAS4A was the second hot corino discovered \citep{Bottinelli2004-iras4a} after IRAS 16293-2422 \citep{Ceccarelli2000-I16293H2CO, Cazaux2003}. 
The iCOMs (interstellar Complex Organic Molecules) emission, the hallmark of hot corinos, was definitively associated with A2 by \cite{Taquet2015}, \cite{Lopez2017-iras4} and \citet{Desimone2017-glyco} \citep[see also][]{Lefloch2018, Quitian-Lara2024-IRAS4survey}. 
Regarding A1, due to the dust opacity in the millimetre, only observations obtained in the centimetre domain revealed methanol emission comparable to that of A2, which makes also A1 an hot corino \citep{Desimone2020-VLA}. 

IRAS4A also has a complex large scale structure, with outflows emanating from each protostar \citep[e.g.,][]{Blake1995, Girart1999-IRAS4Aoutflow,  Lefloch1998a, Choi2005IRAS4Aoutflow, Yildiz2012, Santangelo2015-IRAS4outflows, Desimone2020-I4outflow, Podio2021-jets}. 
Thanks to the recent ALMA/FAUST observations with high-angular resolution ($\sim$50 au), the large and intricate structure of the multiple outflows has been reconstructed \citep{Chahine2024-cav}. 
Relevant to this study, the two sources each have two outflows emanating from them, with a position angle (PA) of 0° and -12° for A1 and 26° and 29° for A2, respectively. 
The northern lobes of both A1 and A2 outflows are red-shifted while the southern ones are blue-shifted. 

Finally, the temperature and density profiles of the large scale envelope surrounding the IRAS4A system (whose radius is about 24000 au) were obtained by \cite{Jorgensen2002}, via a radiative transfer analysis of the continuum emission observed with single-dish telescopes in the millimetre to submillimeter wavelength range. 

\section{OCS binding energy distribution} \label{sec:OCS_BE-computations}

\subsection{Methodology}\label{subsec:BE-methodology}

\subsubsection{Binding energy formalism}\label{subsubsec:BE_definition}

The methodology employed in this work to compute the binding energy distribution follows previous studies from our group \citep{aco_frost, Tinacci2023, Tinacci2022-ammonia, bariosco_h2s, Bariosco2025}. 
The BE, positive for a bound system, is defined as the negative of the interaction energy ($\Delta E$):
\begin{equation}
    \label{eq:BE}
    \mathrm{BE} = -\Delta \mathrm{E} = \mathrm{E}^{\rm iso}_{\rm ads} + \mathrm{E}^{\rm iso}_{\rm grn} - \mathrm{E}_{\rm c},
\end{equation}
\noindent
where $\mathrm{E}_{\rm c}$ is the energy of the adsorbate–grain complex, and $\mathrm{E}^{\rm iso}_{\rm ads}$ and $\mathrm{E}^{\rm iso}_{\rm grn}$ denote the energies of the isolated adsorbate and grain, respectively.

The BE can be decomposed into two contributions:
(i) the pure electronic interaction energy ($\mathrm{BE}_{\rm e}$), corrected if necessary for the basis set superposition error (BSSE), and  
(ii) the geometry deformation energy ($\delta \mathrm{E}_{\rm def}$).  

When the electronic energies are further corrected for the zero-point energy (ZPE) derived from harmonic frequency calculations, the BE becomes the binding enthalpy at 0~K, expressed as:
\begin{equation}
    \label{eq:BE_decompose}
    \mathrm{BE} = \underbrace{\mathrm{BE}_{\rm e} - \delta \mathrm{E}_{\rm def}}_{\mathrm{BE^*}} - \Delta \mathrm{ZPE}.
\end{equation}
In the following, we refer to the ZPE-corrected binding energy as BE, while the uncorrected value is denoted as BE$^*$. 
The ZPE correction, $\Delta \mathrm{ZPE}$, is computed as:
\begin{equation}
    \Delta \mathrm{ZPE} = \mathrm{ZPE}_{\rm c} - \mathrm{ZPE}_{\rm ads}^{\rm iso} - \mathrm{ZPE}_{\rm grn}^{\rm iso},
\end{equation}
where $\mathrm{ZPE}_{\rm c}$ is the ZPE of the adsorbate–grain complex, and $\mathrm{ZPE}^{\rm iso}_{\rm ads}$ and $\mathrm{ZPE}^{\rm iso}_{\rm grn}$ denote the ZPE of the isolated adsorbate and grain, respectively.

\subsubsection{Computational methods}\label{subsubsec:be_definition}

The grain model and initial grain–adsorbate geometries were first optimized at the semi-empirical (SQM) GFN2-xTB level \citep{xtb,GFN2} to obtain preliminary BEs. 
Each preliminary structure was subsequently refined using the ONIOM method \citep{mayhall2010oniom} (QM:SQM), as implemented in the ORCA program (v.~5.0.3) \citep{ORCA}.

In the ONIOM approach, the system is partitioned into two fictitious subsystems: (i) the Real Zone, corresponding to the full system treated at the lower level of theory (SQM); and (ii) the Model Zone, comprising a smaller region that includes sufficient water molecules to capture significant local interactions of the adsorbate, treated at a higher level of theory (QM). 
The ONIOM energy for each structure is evaluated as:
\begin{equation}
  \mathrm{E(ONIOM)} = \mathrm{E_{Real}(SQM)} - \mathrm{E_{Model}(SQM)} + \mathrm{E_{Model}(QM)}.
  \label{eqn:oniom}
\end{equation}
For the high-level QM calculations on the Model Zone, the B97-3c functional \citep{b97_3c} was employed to compute geometries and harmonic frequencies, while GFN2-xTB was retained as the low-level SQM method. 
Atoms outside the Model Zone were kept fixed, and the Model region was embedded electrostatically (default setting). 
Tight convergence criteria were applied to both the self-consistent field (SCF) procedure and geometry optimization, and the integration grid was set to the highest density using the \texttt{!tightscf}, \texttt{!tightopt}, and \texttt{!defgrid3} keywords in ORCA. 
This ensured reliable convergence to minima on the potential energy surface (PES).

Following geometry optimization, single-point energy refinements were carried out using the DLPNO-CCSD(T) method \citep{DLPNO_CCSD(T)_new} within the ONIOM scheme, with GFN2-xTB maintained as the SQM level. 
The aug-cc-pVTZ basis set \citep{kendall1992dunning,woon1993gaussian} was used as the primary basis, while aug-cc-pVTZ/C \citep{pvtz_c} served as the auxiliary basis for the resolution-of-identity approximation. 
DLPNO-CCSD(T) calculations were performed using the TightPNO setup and default SCF convergence thresholds. Finally, the energies of the complexes, $E_{\rm c}$(DLPNO-CCSD(T)), were corrected for basis set superposition error (BSSE) using the counterpoise method \citep{Counterpoise}.

\subsubsection{Water-ice grain model and strategy to compute BEs}\label{subsubsec:BE_strategy}
The initial BE computations were performed as in our previous work \citep{Germain2022}. 
A cluster of 200 water molecules was employed to model interstellar icy grains (see the 200-water grain model at \url{https://aurelegermain.github.io/JSmol_grain/}) \citep{aco_grain}. 
Using the ACO-FROST code \citep{aco_frost}, we generated a spherical grid of 162 points evenly distributed around the cluster. 
An OCS molecule was placed at each grid point, adopting three different initial orientations to maximize the sampling of adsorption geometries, resulting in 486 initial adsorption configurations. 
These starting positions were then projected to distances of 2.5–3.0~\AA{} from the grain surface.

The procedure adopted to compute the BE distribution of OCS on the 200-water cluster involves five steps:
\begin{itemize}
    \item[(\textit{i})] The geometry of the grain model is fixed, and a GFN2-xTB geometry optimization is performed for the adsorbed OCS molecule at each initial position.
    \item[(\textit{ii})] Starting from the structures obtained in step (\textit{i}), all water molecules within a 5~\AA{} region surrounding the relaxed OCS molecule are allowed to relax at the GFN2-xTB level.
    \item[(\textit{iii})] Following the optimization in step (\textit{ii}), a check is performed to confirm that the number of water molecules within the 5~\AA{} region remains unchanged after relaxation. If changes are detected, steps (\textit{ii}) and (\textit{iii}) are repeated until the composition of the 5~\AA{} region stabilizes.
    \item[(\textit{iv})] For the final structures obtained from step (\textit{iii}), two-layer ONIOM (QM:SQM) calculations are carried out. The Model Zone, defined as the 5~\AA{} region, is treated at a high-level quantum mechanical method, while both the Model Zone and the entire cluster (the Real System) are treated at the GFN2-xTB level. The energies are then combined subtractively according to the ONIOM scheme (see Eq. \ref{eqn:oniom}).
    \item[(\textit{v})] During ONIOM geometry optimizations, atoms outside the Model Zone are held fixed. Electrostatic embedding of the Model Zone is enabled via the corresponding keyword. In subsequent frequency calculations (using the harmonic approximation), only the normal modes involving atoms within the Model Zone are considered, while all other nuclei remain fixed. As in steps (\textit{ii}) and (\textit{iii}), it is ensured that the Model Zone retains the same number of water molecules during the optimization. The energy of the isolated grain surface ($E_{\rm grn}^{\rm iso}$) is computed following our previous “TPD” approach \citep{Tinacci2023}, which closely mimics the experimental temperature-programmed (TPD) desorption process by relaxing the QM system after removing the adsorbed OCS molecule.
\end{itemize}

\subsubsection{Procedure to obtain unique BE sites}\label{subsubsec:unique-BEsites}

In several cases, different initial orientations of the OCS molecule converge to the same final optimized structure. 
To prevent redundancies that could distort the final BE distribution, a pruning procedure was developed. 
This step is essential to eliminate duplicate BE sites arising from slightly different initial geometries that relax to the same minimum on the PES.

Redundant sites are identified through the following steps:
\begin{itemize}
    \item[(\textit{i})] All pairs of Model Zone structures with the same number of atoms are selected for comparison.
    \item[(\textit{ii})] For each pair, the structures are aligned spatially, and the absolute energy difference is calculated as:
    \[
    \rm{|\Delta E_{Cij}| = |E_{Ci} - E_{Cj}|,}
    \]
    where $E_{Ci}$ and $E_{Cj}$ denote the energies of the $i$-th and $j$-th complex structures, respectively.
    \item[(\textit{iii})] Two structures are considered identical if their root-mean-square deviation (RMSD) is below 0.5~\AA{} and if $\rm|\Delta E_{Cij}|$ is less than 1.0~kJ/mol.
\end{itemize}

The thresholds of 0.5~\AA{} for RMSD and 1.0~kJ/mol for energy difference were established after extensive visual inspection of potentially redundant BE sites. 
For a detailed analysis on the procedure refer to \citet{bariosco_h2s}.

\subsubsection{The desorption rate prefactor}\label{subsubsec:prefact}

As pointed out by previous articles \citep{minissale2022thermal, ferrero2022acetaldehyde, Tinacci2023, Ligterink2023, Ceccarelli2023-PP7, Pantaleone2025-prefactor}, the prefactor is an extremely important parameter to compute the desorption rate of a species (see $k_{des}$ in Eq.~\ref{eq:k-des-ads}). 
We here adopted the \citet{tait2005n} formula for linear molecules ($I_x = 0$), which the recent study by \cite{Pantaleone2025-prefactor} has shown to be a good approximation of the accurate QM calculations:
\begin{equation}
\label{eq:apprx_pre_exponential}
    \nu_{des}(\mathrm{T}) = \frac{\mathrm{k_B T}}{\mathrm{h}} \Biggl(\frac{2 \pi \mathrm{~m ~k_B T}}{\mathrm{h^2}} \Biggr) \mathrm{~A} \frac{\sqrt{\pi}}{\sigma \mathrm{h}^3} \bigl(8 \pi^2 \mathrm{k_B T} \bigr)^{\frac{3}{2}} \sqrt{\mathrm{I_y ~ I_z}} \; {\rm ~s^{-1}},
\end{equation}
where $\mathrm{k_B}$ is the Boltzmann constant, m the mass of the molecule, h the Planck constant,  A is the surface area per adsorbed molecules (usually  assumed to be $10^{-19} $m$^2$), $\mathrm{I_i}$ is the adsorbate principal moment of inertia in the \textit{i-esimal} direction, and $\sigma$ is the symmetry adsorbate rotation factor.  
For OCS, the principal moments of inertia are 83.6, 83.6 a.m.u.$~$\AA$^2$, $\sigma$ is 1 and m is 60.07 a.m.u..

\subsection{Results}\label{subsec:BE-results}
After the pruning procedure, 320 redundant sites are identified and discarded from the distribution, resulting in 166 unique sites. 
All the obtained structures are PES minima, \textit{i.e.} imaginary frequency are not present. 
The final distribution is reported in Fig. \ref{fig:dlpno_distribution}. 
The number of bins and their width are obtained following the Freedman Diaconis estimator \citep{freedman1981histogram}. 
As shown in Fig. \ref{fig:dlpno_distribution}, the histogram is well-fitted by a Gaussian distribution:
\begin{figure}
\includegraphics[width=1.0\columnwidth]{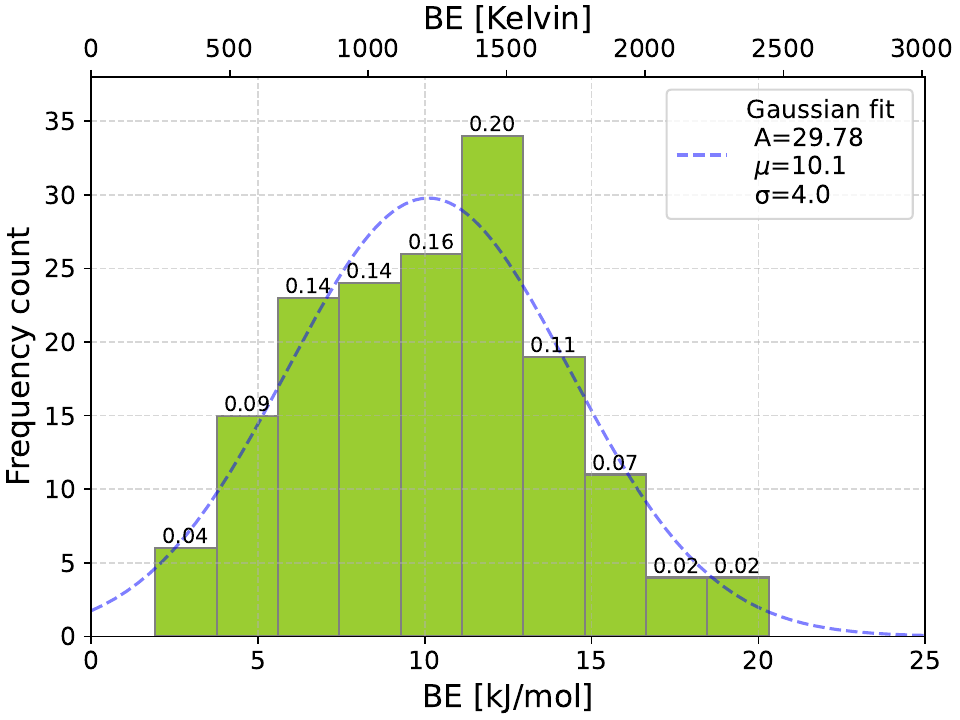}
    \caption{BSSE corrected BE distribution of OCS on a model icy cluster at DLPNO-CCSD(T) level. Structures and ZPEs are calculated at ONIOM(B97-3c:GFN2-XTB) level. The dashed blue curve is the f$_\mathrm{Gauss}$(\texttt{hist}(BE), $\sigma$, $\mu$) not normalized Gaussian best fit function for the histogram.}
    \label{fig:dlpno_distribution}
\end{figure}

\begin{table}
    \centering
    \begin{tabular}{c c c c }
    \hline
         Mean BE     & Pre-factor ($\nu_{des} (T_{peak})$) & $T_{peak}$ & Ice fraction\\
         \hline
           2.8  (341) &  1.4 $\times$ 10$^{14}$ & 10 & 0.04 \\
           4.7  (561) &  7.5 $\times$ 10$^{14}$ & 16 & 0.09 \\
           6.5  (782) &  2.2 $\times$ 10$^{15}$ & 21 & 0.14 \\
           8.4  (1003) & 4.6 $\times$ 10$^{15}$ & 26 & 0.14 \\
           10.2 (1223) & 8.9 $\times$ 10$^{15}$ & 32 & 0.16 \\
           12.1 (1444) & 1.5 $\times$ 10$^{16}$ & 37 & 0.20 \\
           13.9 (1665) & 2.3 $\times$ 10$^{16}$ & 42 & 0.11 \\
           15.7 (1885) & 3.5 $\times$ 10$^{16}$ & 47 & 0.07 \\
           17.6 (2106) & 5   $\times$ 10$^{16}$ & 52 & 0.02 \\
           19.4 (2327) & 6.7 $\times$ 10$^{16}$ & 56 & 0.02 \\
        \hline
    \end{tabular}
    \caption{OCS BE distribution.
    For each bin of the histogram (Fig. \ref{fig:dlpno_distribution} the columns report the mean BE value, pre-factor $\nu_{des}$ evaluated at the T$_{peak}$, T$_{peak}$ and the fraction of ice, respectively.)
    The energy values are in kJ/mol and reported in parenthesis in Kelvin (K). 
    The pre-factor is in s$^{-1}$. 
    The bin width is 1.84 kJ/mol (221 K).}
    \label{tab:population_distrib}
\end{table}
\begin{equation}
    \mathrm{G (BE) =A \exp\left({-\frac{(BE-\mu)^2}{2\sigma^2}}\right)},
    \label{eqn:gaussian}
\end{equation}
where A is equal to 29.8, which represents the maximum height of the fitted Gaussian curve, the mean value of the BE $\mu$ is 10.1 kJ/mol (1215 K) and the standard deviation is 4.0 kJ/mol (481 K). 
The maximum value found is 20.3 kJ/mol (2436 K), while the lower is 1.9 kJ/mol (228 K). 
Table \ref{tab:population_distrib} reports the values of the various bins, for an easy incorporation into astrochemical models. 
The prefactor $\nu_{des}$ was computed at the desorption peak (T$_{peak}$), solving numerically the Polany-Wigner equation \citep{polanyi1997bildung}, as done in a previous work by our group \citep{bariosco_h2s}.

\subsection{Comparison with previous results}\label{subsubsec:BE-comparison}

Table \ref{tab:be_comparison} summarizes the BEs determined through experimental measurements, computational studies, and those adopted in major astrochemical databases. While systematic uncertainties associated with the chosen theoretical level and model assumptions cannot be completely ruled out, the present computational setup provides the most accurate and consistent ab initio description currently achievable for systems of this size and complexity. Overall, our newly computed BE distribution reveals a range of values not previously reported in the literature. The mean BE in this work (1215 K) is approximately 800–1600 K lower than the values reported in earlier studies listed in Table \ref{tab:be_comparison}.

\begin{table}
    \begin{tabular}{lcl}
    \hline
    Reference & BE in K (ASW) & Result \\
    \hline
    \citet{Wakelam2017BE}            & 2100  & Computational \\
    \citet{Das2018}                  & 1571  & Computational \\
    \multirow{3}{*}{\citet{Perrero2022-SBE}} &  Min: 1286 & \multirow{3}{*}{Computational} \\
                                     & Max: 2861 &  \\
                                     & $\mathbf{\mu}$: 2083\textsuperscript{\emph{a}} & \\
    \citet{Penteado2017}             & 2325  & Experimental \\
    \citet{Wakelam2015}              & 2400  & Database (KIDA) \\
    \citet{McElroy2013}              & 2888  & Database (UMIST) \\
    \multirow{3}{*}{This work}       & Min: 229 & \multirow{3}{*}{Computational} \\
                                     & Max: 2442 &  \\
                                     & $\mathbf{\mu}$: 1215 & \\
    \hline
    \end{tabular}
    \caption{Comparison of the OCS BE quoted in other works (computational and experimental) and databases with the BE distribution of the present work.
    Notes: \textsuperscript{\emph{a}} \citet{Perrero2022-SBE} reported a range of values and not a statistical distribution as done in this work.}
    \label{tab:be_comparison}
\end{table}

\begin{table*}
    \centering
    \begin{tabular}{c|ccccc|cccc}
        \hline
        \multirow{2}{*}{Transition} & $\nu^{(a)}$ & E$_{\rm{up}}^{(a)}$ & A$_{\rm{ij}}^{(a)}$ & S$\mu^{2 (a)}$ & dV & FWHM FoV & Beam (PA) & Chan. rms & \multirow{2}{*}{Weighting mode}\\
        & (MHz) & (K) & (10$^{-5}$ s$^{-1}$) & (D$^2$) & (km/s) & $\arcsec$ & $\arcsec$ $\times$ $\arcsec$ (°) & (mJy/beam) & \\
        \hline
        \multirow{2}{*}{OCS(19--18)} 
            & \multirow{2}{*}{231060.983} & \multirow{2}{*}{110.9} & \multirow{2}{*}{3.57} & \multirow{2}{*}{9.72} & \multirow{2}{*}{0.183} &  \multirow{2}{*}{27.3} &0.307 $\times$ 0.204 ($-$7) & 0.91 & Natural\\ 
         & & & & & & & 0.174 $\times$ 0.110 ($+$180) & 1.31 & Robust\\ 
        \hline
\end{tabular}
    \caption{Spectroscopic \citep[$^{(a)}$][from the CDMS database \citep{Muller2005}]{Golubiatnikov2005-OCS} and observational parameters of the OCS data. The resulting beams and channel rms are for the natural and robust weighting, respectively.}
    \label{tab:ocs-parameters}
\end{table*}

The BE calculated by \citet{Das2018}, employing a water tetramer to model the ice surface, closely matches our mean value, whereas the BE estimated by \citet{Wakelam2017BE}, based on a single water molecule and corrected via an empirical factor, lies at the upper end of our range. 
Notably, the minimum BE reported by \citet{Perrero2022-SBE} aligns with the mean value obtained in this study. 
This agreement may reflect the limited surface sampling in their work, which considered only eight binding sites selected through chemical intuition, potentially introducing bias. 
Furthermore, in \citet{Perrero2022-SBE}, the ZPE and CCSD(T) corrections were applied solely to a subset of crystalline structures and then extrapolated to amorphous systems. 
As highlighted by \citet{bariosco_h2s}, such an approach can substantially affect the accuracy of the final BE values.

The experimental BE reported by \citet{Penteado2017} in \autoref{tab:be_comparison} was derived from the $T_{peak}$ values measured by \citet{Collings2004}, using a linear correlation where water serves as the reference species. 
This experimental estimate lies near the upper boundary of our computed distribution. 

Finally, the BE values listed in the KIDA and UMIST databases are 2400 and 2888 K, respectively. 
While the former corresponds to the high-energy tail of our distribution, the latter exceeds our maximum computed value by more than 400 K.

\section{Observations} \label{sec:observations-results}

\begin{figure*}
    \centering
    \includegraphics[width=1\linewidth]{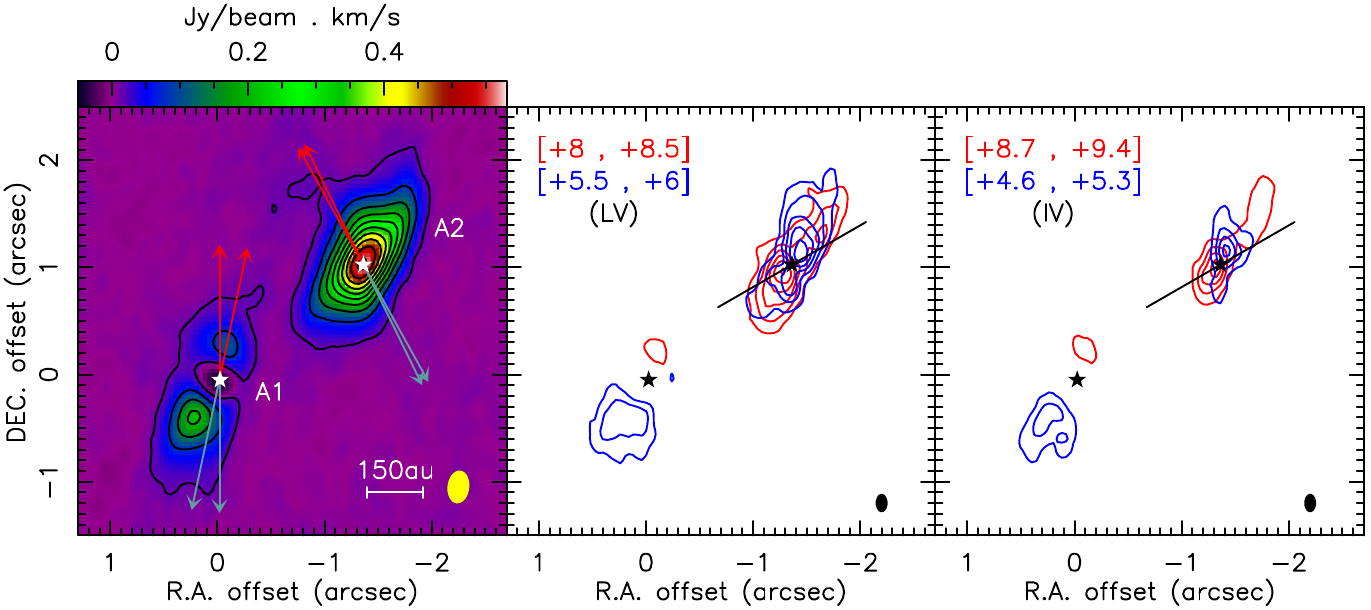}
    \caption{\textit{Left panel:} Moment 0 of OCS(19--18) integrated between $-$0.15 km/s and +14.64 km/s, where the sensibility is optimised. 
    Contours start at 5$\sigma$ ($\sigma$ = 3.03\,mJy/beam\,km/s) and increase by steps of 20$\sigma$. 
    Both  position of the two protostars A1 and A2, is represented by white stars.
    Note that for A2 the position is taken to be the peak of OCS emission (see text).
    Red and blue arrows represent the red-shifted and blue-shifted outflow directions, respectively \citep{Chahine2024-cav}. 
    The synthesised beam is the yellow ellipse in the bottom right corner of the map.
    \textit{Middle and right panels:} Moment 0 maps of OCS(19--18) integrated over different velocity intervals: $\pm[1;1.5]$ km/s (LV) and $\pm[1.7;2.4]$ km/s (IV) with respect to V$\rm{_{lsr}}$ = 7 km/s. 
    Contours start at 5$\sigma$ ($\sigma$ = 0.62\,mJy/beam\,km/s and 0.70\,mJy/beam\,km/s for LV and IV panels, respectively) and increase by steps of 10$\sigma$.
    Note that, in this case, we optimised the resolution (see text).
    Both  position of the two protostars A1 and A2 are represented by black stars.
    Black lines represent the direction of the envelope where OCS is thermally sublimated and used for the model analysis (see text).
    The synthesised beam is the black ellipse in the bottom right corner of each map.}
    \label{fig:ocs-mom0}
\end{figure*}

\subsection{Description of the observations} \label{subsec:obs}
The IRAS\,4A system was observed at 1.2\,mm with ALMA (Cycle 6), between December 2018 and September 2019, as part of the ALMA Large Program FAUST \citep[proposal 2018.1.01205.L, PI: S. Yamamoto;][]{Codella2021}. 
The observations were performed in Band 6 using the 12-m array (C43-3 and C43-6 configurations, including 46 and 50 antennas, respectively). 
The baselines for the 12-m array ranged between 15.1\,m and 2.5\,km, probing angular scales from 0$\farcs$10 ($\sim$30\,au) to 8$\farcs$2 ($\sim$2400\,au).

The observations were centered at R.A. (J2000)\,=\, 03$^{\rm{h}}29^{\rm{m}}10^{\rm{s}}$.539, Dec. (J2000)$ \,= 31^{\circ}13'30 \arcsec 92$ (i.e. A1 position) and the systemic velocity was set to V$_{\rm{lsr}}$\,=\,7\,km\,s$^{-1}$ (see Introduction). 
Several spectral windows (spw) were placed within the spectral range 216–234 GHz.

Data calibration (including self-calibration) was performed using the ALMA calibration pipeline with the Common Astronomy Software Applications package (CASA\footnote{\url{https://casa.nrao.edu/}}, \citealp{CasaTeam2022}).
More details on calibration can be found in \cite{Chahine2024-cav} which used the same data set.

In this work, we focus on the narrow spw covering the OCS(19-18) line observed with the C43-6 configuration to obtain a better resolution to distinguish the envelope of the two sources.
The spw is centered at 231060.983 MHz on the OCS(19-18) line with a total bandwidth of 58.6 MHz ($\sim$76.0\,km/s) and a channel width of 141 kHz ($\sim$0.183\,km/s).
The resulting parameters are obtained by cleaning with natural and robust weighting, depending on whether we needed better sensitivity or spatial resolution, respectively.
The spectroscopic parameters, along with beam size and rms are described in Table \ref{tab:ocs-parameters}.


\subsection{Results} \label{subsec:obs-results}

\begin{figure*}
    \centering
    \begin{tabular}{cccc}
        \multicolumn{2}{c}{\includegraphics[width=0.39\linewidth]{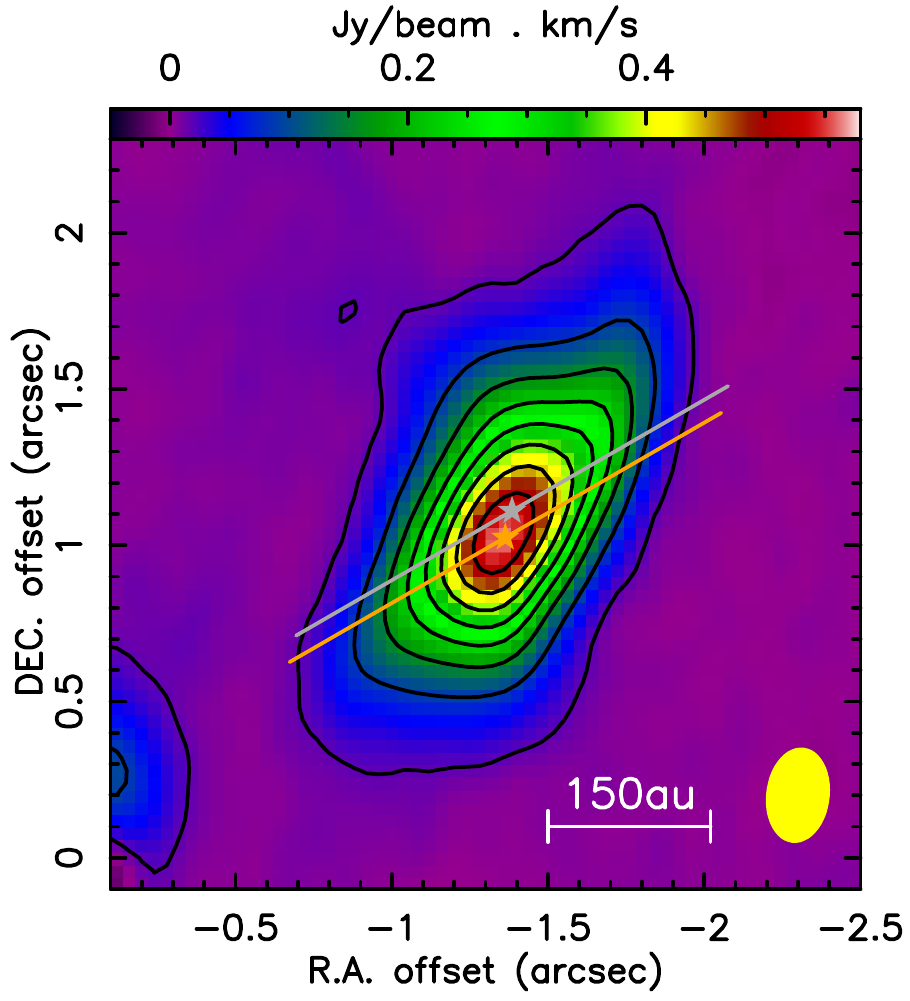}} & \includegraphics[width=0.22\linewidth]{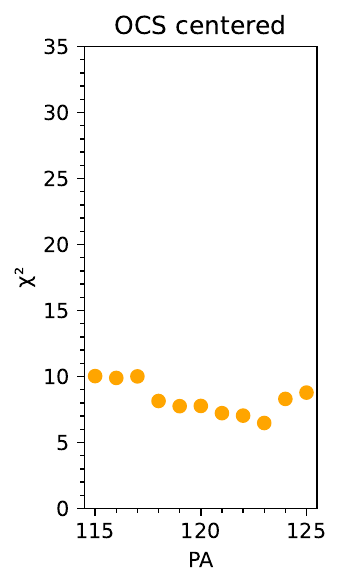} & \includegraphics[width=0.22\linewidth]{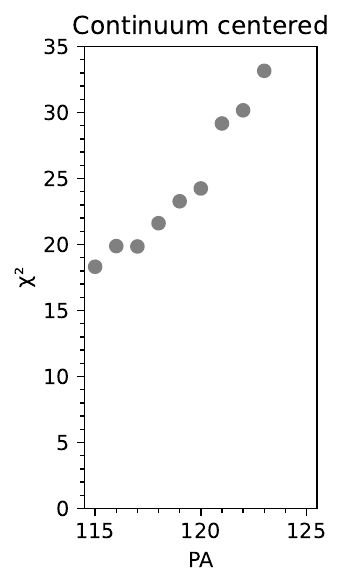} \\
        \multicolumn{2}{c}{\includegraphics[width=0.48\linewidth]{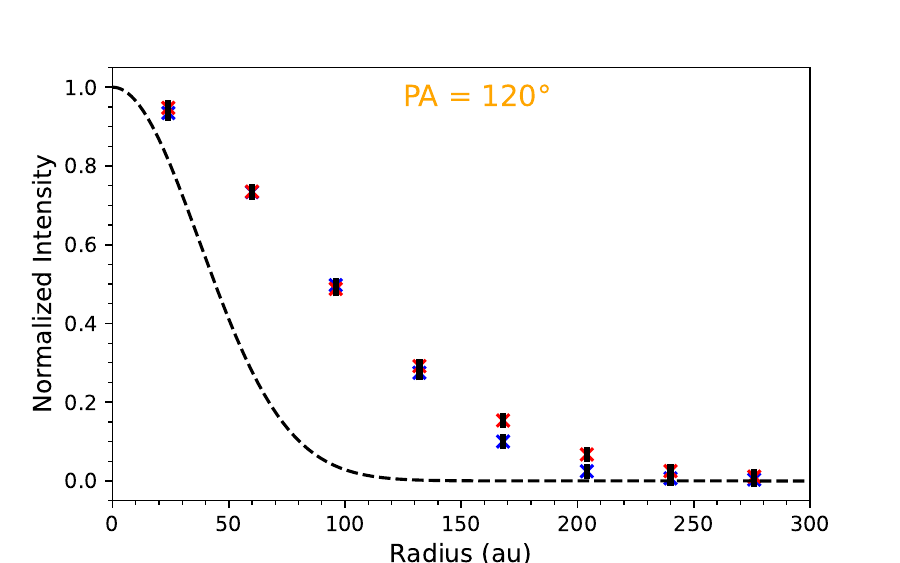}} & \multicolumn{2}{c}{\includegraphics[width=0.48\linewidth]{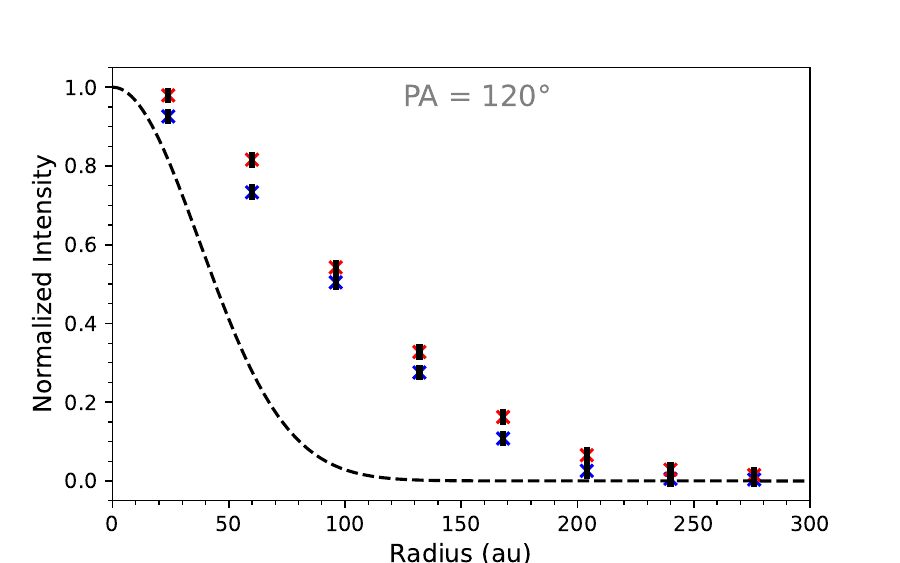}} \\
    \end{tabular}
    \caption{\textit{Top left panel:}
    Moment 0 of the OCS line intensity around A2. 
    The intensity is integrated between +2.74 km/s and +14.17 km/s. 
    Contours start at 5$\sigma$ ($\sigma$ = 3.5\,mJy/beam\,km/s) and increase by steps of 20$\sigma$. 
    The blue-shifted emission corresponds to the west side of the star, while the red-shifted emission to the east part (see Fig. \ref{fig:ocs-mom0}). 
    The grey star marks the continuum emission peak while the orange star the OCS emission peak. 
    The grey and orange lines show the PA = 120°, centred on the two emission peaks, respectively.
    The synthesised beam is represented in yellow in the bottom right corner of the map.
    \textit{Top centre and right panels:}
    $\chi^2$ of the difference between blue- and red-shifted line intensity varying the PA between 115 and 125°, for the centre set on the OCS line (orange) and dust continuum (grey) emission peaks, respectively. 
    \textit{Bottom panels:}
    Normalized OCS line intensity as a function of the distance from the centre set on the OCS line (left panel) and dust continuum (right panel) emission peaks with PA equal to 120°.    
    The red and blue crosses represent the OCS line intensity toward the red- and blue-shifted emission, respectively. 
    The line intensity error bars, computed as 3$\sigma$ ($\sigma$ = 3.5\,mJy/beam\,km/s), are the solid black lines, mostly masked by the crosses. 
    The black dashed line shows the synthesised beam profile.}
    \label{fig:pa-dif-pos}
\end{figure*}

\subsubsection{OCS map} \label{subsubsec:ocs-map}
Figure \ref{fig:ocs-mom0} left panel shows the moment 0 emission map of the OCS(19--18) line emission integrated between $-$0.15 km/s and +14.64 km/s around A1 and A2 to cover the whole line profile. 
The middle and right panels of Fig. \ref{fig:ocs-mom0} show the emission at different velocity intervals, the low velocity (LV) at $\pm[1;1.5]$ km/s and intermediate velocity (IV) at $\pm[1.7;2.4]$ km/s with respect to the systemic velocity V$\rm{_{lsr}}$ = 7 km/s. 
The moment 0 map was cleaned using a natural weighting to obtain a better sensitivity, while the two channel maps were cleaned with a robust weighting, which allows a better spatial resolution with respect to the moment 0 map. 
This allows us to better disentangle the emission originating from the envelope from that of the cavities. 
Especially for A2, the LV channel map shows that the emission arises from both the envelope and the cavities, whereas the IV map better distinguishes each component in the direction where the envelope dominates the emission (see \S ~\ref{subsubsec:ocs-a2} for more details).

Around A1, the OCS line absorption is clearly visible and it is due to the high dust column density close to the protostar. 
In this source, the OCS emission is only visible at the base of the cavities opened by the two outflows from A1, along the directions previously identified by \cite{Chahine2024-cav}, and shown in Fig. \ref{fig:ocs-mom0} by red and blue arrows, for the red- and blue- shifted lobes, respectively. 
Around A2, the dust emission is less optically thick, allowing the OCS emission from the two cavities and the inner envelope to be detected over a region of $\sim 450\times250$ au (based on the moment 0 map). 

In the following, we will only focus on the emission coming from the inner envelope around A1 and A2.
Our goal is to identify the regions where OCS is gaseous because of its thermal sublimation.
Since at the cavity bases frozen OCS may be released from the grain mantles by non-thermal processes, such as sputtering and shattering by mild shocks, we will exclude those regions from our analysis.
Finally, because of the heavy dust absorption, this will not be possible to do towards A1 and we will restrict this analysis to A2 only.

\subsubsection{OCS emission in A1} \label{subsubsec:ocs-a1}
Because of the dust absorption around A1, the OCS emission near the centre of the protostar is hidden. 
In order to constrain the sizes of the inner envelope region where OCS is thermally sublimated, we used the known position of the cavities to estimate the PA of the envelope, that we suppose to be roughly perpendicular to the cavities. 
Using the cavities' PAs estimated by \cite{Chahine2024-cav}, the PA of the inner envelope is then 70$\pm$5°.
Along this axis, the OCS line emission is totally absorbed up to a radius of $\sim$140 au. 
Therefore important we cannot extract meaningful quantitative information from the OCS emission around A1.

\subsubsection{OCS emission in A2} \label{subsubsec:ocs-a2}
The OCS emission around A2 is more complex than in A1, as it results from the contribution of both the cavities excavated by the two outflows emanating from A2 and its inner envelope, where OCS is thermally sublimated.
In order to constrain where the OCS emission is dominated by the inner envelope, once again we used the results by \cite{Chahine2024-cav} to constrain its major axis. 
By using the direction perpendicular to that of the cavities, the inner envelope is expected to have a PA close to 120°, represented by the black line in Fig. \ref{fig:ocs-mom0}. 
In addition, in order to better refine the emission originating (mostly) from the inner envelope, we used the OCS line emission as a function of the distance from the presumed centre of the inner envelope of the red- (south) and blue- shifted (north) emitting regions.

When the emission originates in the inner envelope, the red- and blue- shifted emitting regions identify the receding and approaching regions of the rotating/infalling gas so that the two profiles should coincide, while an asymmetry would point to contamination from the (asymmetric) outflow cavities. 
This emission is visible in the LV and IV maps in Fig. \ref{fig:ocs-mom0}, with an asymmetric extended emission, especially in the red-shift of the IV panel.
We, therefore, minimized the difference of the red- and blue- shifted line emission profiles varying the PA between 115 and 125°, and centring the inner envelope either on the continuum peak or OCS line emission peak in the moment 0 map, respectively.
To this end, we used a $\chi^2$ method, where $\chi^2$ is the the sum of the squares of the difference between the blue- and red- shifted normalized intensity emission at the same distance from the centre, normalized for the square of the signal uncertainties.

The results are shown in Fig. \ref{fig:pa-dif-pos}.
The lowest $\chi^2$ is obtained when the centre is set at the OCS emission peak and for a PA between 118 and 123°.
The bottom panels of Fig. \ref{fig:pa-dif-pos} presents the OCS line intensity emission at a PA of 120° extracted along the blue- and red-shifted sides of the envelope.

\section{Model of the OCS line emission profile} \label{sec:model}

As discussed in \S ~\ref{sec:intro}, our goal is to determine the luminosity of an embedded protostar in a binary system using the OCS observed line intensity profile and comparing it with the theoretical one, obtained using the OCS BE distribution, described in \S ~\ref{subsec:BE-results}.
In the specific case of the IRAS4A binary system considered here, we could only obtain the profile around A2, as the OCS line emission is completely absorbed around A1 (Sec. \ref{subsec:obs-results}).
Therefore, we applied the method to A2 to derive its luminosity and then constrained the A1 luminosity from the measured total A1+A2 luminosity, as discussed in the next Section.

In this section, we describe the theoretical model.
The theoretical OCS line intensity profile around A2 depends on the adopted physical structure (temperature and density profiles; \S ~\ref{subsec:gas-profile}), OCS abundance profile (\S ~\ref{subsec:abu-profile}) and radiative transfer (\S ~\ref{subsec:model-line-intensity}), described separately in the following.

\begin{figure*}
    \centering
    \begin{tabular}{cc}
         \includegraphics[width=0.5\linewidth]{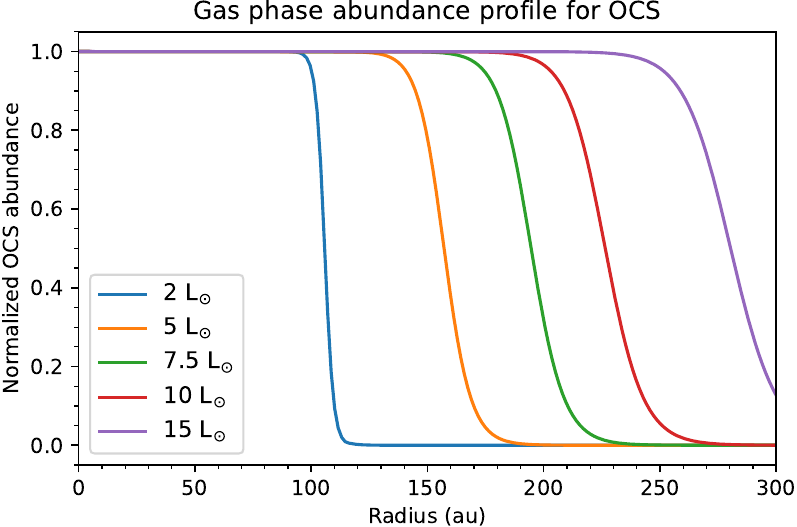}
         \includegraphics[width=0.5\linewidth]{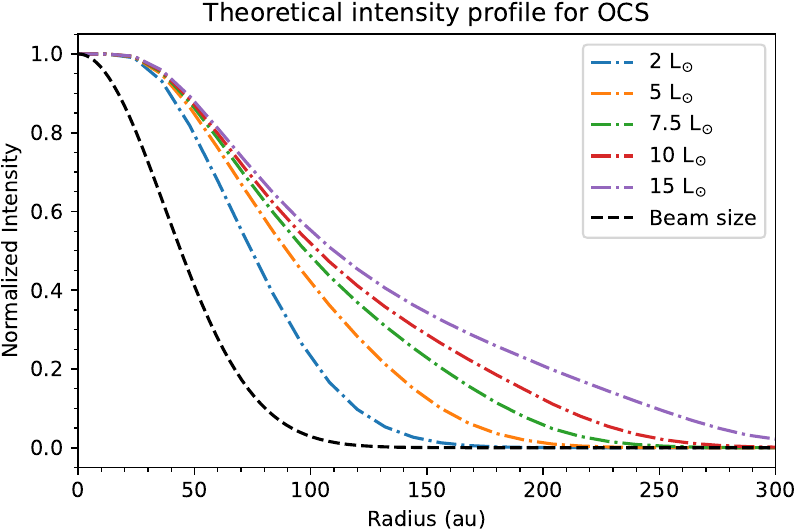} 
    \end{tabular}
    \caption{\textit{Left panel:} Normalised OCS gas abundance profile towards A2 for five luminosities: 2 (blue), 5 (orange), 7.5 (green), 10 (red) and 15 (purple) L$_\odot$.
    \textit{Right panel:} Theoretical intensity profile depending on the luminosity value. The
    dashed line shows the synthesised beam profile of 0$\farcs$25.}
    \label{fig:OCS_abundance-intensity}
\end{figure*}
\subsection{Gas density and temperature profiles} \label{subsec:gas-profile}

We used the density and temperature profiles of the IRAS4A spherical envelope derived by \cite{Jorgensen2002} via a radiative transfer modelling of a large set of dust continuum observations.
The derived profiles cover a radius between 24 and 24000 au and were obtained assuming a single source IRAS4A with a luminosity of 6 L$_\odot$ and a distance of 220 pc. 
While the absolute temperature is mostly set by the protostar luminosity, the modelled density profile depends very little on it (see below).
We discuss the adopted assumptions and their limits in the following (see also \S ~\ref{subsubsec:disc-temp-profile}).

\noindent
\textit{Density profile:}
We assumed the power law derived by \cite{Jorgensen2002} at the scales of the present observations ($\sim$20--300 au):
\begin{equation}\label{eq:dens-profile}
    n_{gas}(r)  ~=~ 4\times10^{8}  ~\left( \frac{r}{100 ~{\rm au}} \right)^{-1.8} ~{\rm cm^{-3}}.
\end{equation}
We only compare the line intensity profile normalised to the maximum, so that what matters is the power law index.
In \S ~\ref{subsubsec:disc-dens-profile} we discuss how the adopted index impacts the results and also show that the observed OCS line intensity profile provides a strict constraint to it.

\noindent
\textit{Temperature profile:}
In general, the temperature profile is described by two power laws corresponding to the dust optically thick and thin regions, respectively \citep[e.g.][]{Adams1986-IRspectra}.
Here we assumed the temperature profile derived by \cite{Jorgensen2002} at the scales of the present observations ($\sim$20--300 au), after scaling it for the larger luminosity.
To this end, we first fitted the \cite{Jorgensen2002} temperature profile with two power laws and then scaled the absolute temperature for the higher luminosity, using the theoretical dependence on luminosity to the 1/4 power \citep[e.g.][]{Adams1985-IRspectra, Ceccarelli2000-I16239structure}.
It holds:
\begin{eqnarray}\label{eq:temp-profile}
    {\rm 20 \leq r \leq 115 ~au:} & T_{dust}(r) ~=~ 31 ~
              \left(\frac{L_{A2}}{L_{\odot}}\right)^{1/4} 
              \left(\frac{r}{115 ~{\rm au}} \right) ^{-1} ~{\rm K}\\
    {\rm 115 < r \leq 300 ~au:} & T_{dust}(r) ~=~ 31 ~
                  \left(\frac{L_{A2}}{L_{\odot}}\right)^{1/4} 
              \left(\frac{r}{115 ~{\rm au}} \right) ^{-0.5} ~{\rm K}
\end{eqnarray}
We verified that our assumption of scaling the absolute temperature with L$^{1/4}$ holds, by analysing the theoretical profiles obtained by \cite{Jorgensen2002} for sources with different luminosities.
In addition, in \S ~\ref{subsubsec:disc-temp-profile} we discuss the impact of a possible different dependence.
Finally, note that, given the involved large density ($\geq 10^8$ cm$^{-3}$) of the A2 inner envelope, we assumed that the dust and gas are thermally coupled, i.e., $\rm T_{dust}=T_{gas}$.

\subsection{OCS gaseous abundance profile} \label{subsec:abu-profile}

So far, OCS is the only S-bearing species firmly detected in interstellar ices \citep[e.g.][]{Boogert2022, McClure2023-JWSTices} and its abundance is measured to be $\sim 10^{-3}$ times that of solid water, namely about $10^{-7}$ with respect to H$_2$.
Based on theoretical quantum chemical computations, OCS is inefficiently formed in the gas-phase \citep{Loison2012-OCS}, so it is reasonable to assume that its gaseous abundance is dominated by the desorption of frozen OCS and its adsorption onto the grain mantles.
The OCS freezing timescale in the inner envelope, where the H$_2$ density is 50 -- 1 $\times 10^8$ cm$^{-3}$, is lower than about 50 yr.
Since the estimated age of IRAS4A is $\sim 10^{4-5}$ yr \citep{Maret2002, Kristensen2018-ProtostarsLifetime}, we assumed a steady state solution, given by equating the desorption and adsorption rates of OCS from the ice to the gas phases.
This assumption would be invalid if IRAS4 has bursts with timescales lower than about 60 yr.
Recent observations of the dust continuum, which probes the dust temperature, show variability lower than a few percent, which implies a luminosity burst of less than 50\% \citep{Johnstone2018-variability, lee2021-variability, Mairs2024-variability}.
We conclude that our assumption of steady state is correct for IRAS4A.

As discussed in Sec. \ref{sec:OCS_BE-computations}, OCS molecules are bound to the water-rich icy surface with a distribution of BEs, which covers a range of 2.9--19.4 kJ/mol ($\sim$340--2330 K).
Assuming the steady state means that OCS molecules are constantly adsorbed and desorbed until an equilibrium is reached.
In practice, a molecule in a low BE site will desorb into the gas-phase and it will then be adsorbed in one of the BE sites with a probability equal to the distribution shown in Fig. \ref{fig:dlpno_distribution} and reported in Table \ref{tab:population_distrib}.
With time, therefore, the low BE sites will be depopulated in favour of the high BE ones, in those where the BE is large enough that the desorption becomes impossible.
When following these considerations, the abundance of gaseous OCS is easy to estimate, because it is governed by the BE bin with the highest value.
Note that this would not apply to species with a large enough abundance (with respect to the water ice), for which (some) large BE sites could be saturated (e.g. CH$_3$OH). 
In the case of OCS, where the abundance with respect to the ice-water is $\sim 10^{-3}$, this never occurs (when considering its BE distribution).

Therefore, in order to compute the OCS abundance profile we solved the following equations:    
\begin{eqnarray}\label{eq:chem-steadystate}
    k_{des}~ {\rm x_{ice}(OCS)} ~=~ k_{ads}~ {\rm x_{gas}(OCS)}\\
    {\rm x_{ice}(OCS)} ~+~ {\rm x_{gas}(OCS)} ~=~ {\rm x_{tot}(OCS)} 
\end{eqnarray}
where ${\rm x_{ice}(OCS)}$, ${\rm x_{gas}(OCS)}$ and ${\rm x_{tot}(OCS)}$ are the OCS iced, gaseous and total abundances, respectively.
The computed line intensity profile being normalised to unity, ${\rm x_{tot}(OCS)}$ is taken equal to 1.
$k_{des}$ and $k_{ads}$ are the desorption and adsorption rates, given by:
\begin{eqnarray}\label{eq:k-des-ads}
    k_{des} = \nu_{des} ~exp\left[{-\frac{BE}{T_{dust}}}\right]\\
     k_{ads} = S ~\pi ~a_{dust}^2 ~n_{dust} ~{\rm v}_{th}
\end{eqnarray}
where $\nu_{des}=6.7\times10^{16}$ s$^{-1}$ and $BE$=19.4 kJ/mol (2327 K) are the pre-factor and BE of the highest energy bin of the OCS BE distribution (Tab. \ref{tab:population_distrib}), respectively\footnote{BE is an energy that can be expressed as K and is related to the desorption temperature as, e.g., described in \cite{Ceccarelli2023-PP7}.};
$S$ is the sticking coefficient, assumed to be 1;
$a_{dust}$ is the average radius of the dust grains, equal to 0.1 $\mu$m; 
${\rm v}_{th}$ is the thermal velocity, defined as $\sqrt{\left(2 ~k_B ~T_{gas} \right) / \left(m_{OCS}\right)}$, with $m_{OCS}$ the OCS mass, equal to 60 $m_H$.

Figure \ref{fig:OCS_abundance-intensity}, left panel, shows the resulting gaseous OCS abundance profile, normalised to 1, as a function of the envelope radius, for five luminosities: 2, 5, 7.5, 10 and 15 L$_\odot$.
As expected, the larger the luminosity the larger the regions where OCS is in the gas-phase.
The drop in the abundance is sharp, due to the exponential in the desorption rate.

\begin{figure*}
    \centering
    \begin{tabular}{c}
        \includegraphics[width=0.48\linewidth]{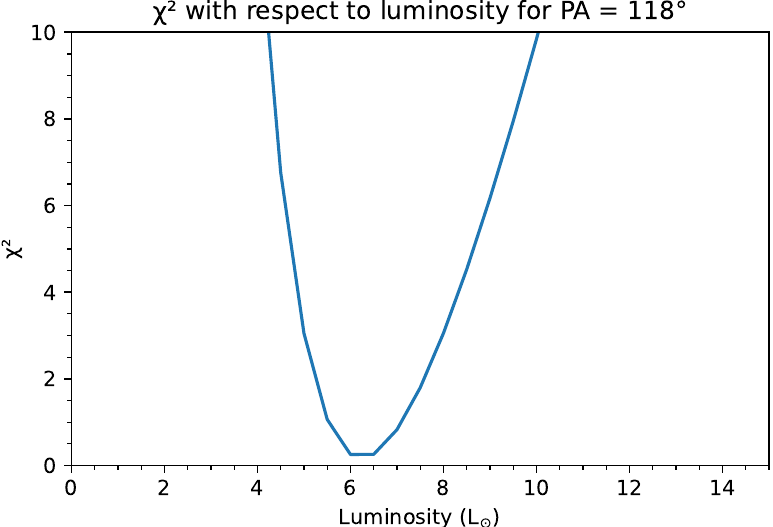} 
        \includegraphics[width=0.48\linewidth]{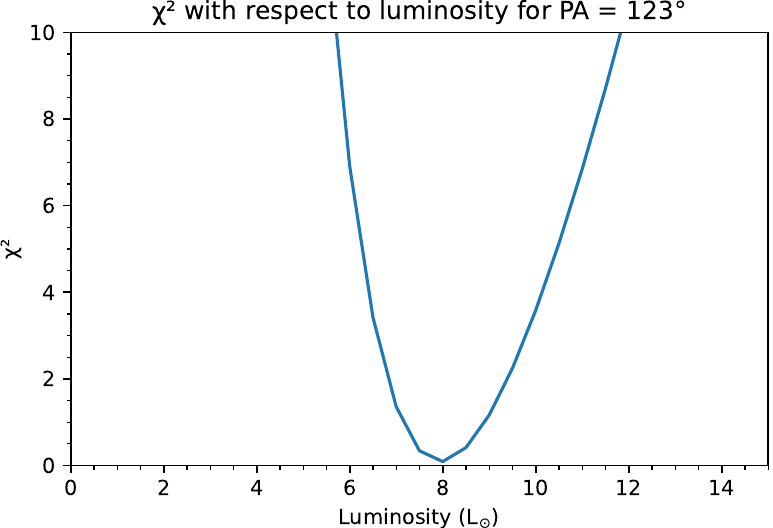} \\
        \includegraphics[width=0.48\linewidth]{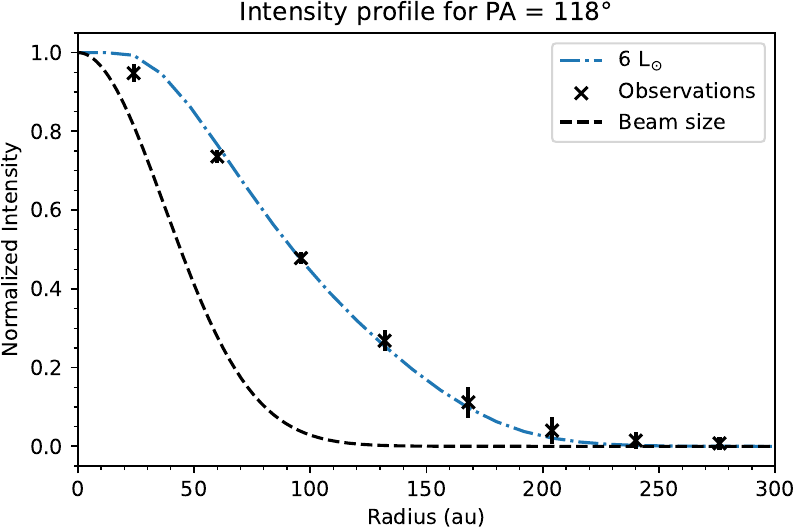} 
        \includegraphics[width=0.48\linewidth]{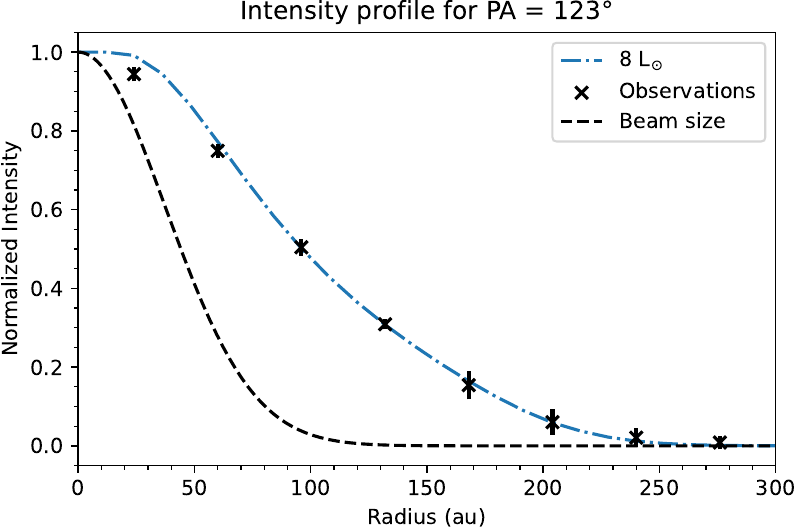} 
    \end{tabular}
    \caption{\textit{Top panels:}
    Normalised $\chi^2$ as a function of the A2 luminosity, for PA=118° (left) and 123° (right),  where the two innermost data points of the observations are not considered (see text). 
    \textit{Bottom panels:}
    Normalised line intensity profile towards A2.
    The blue dashed lines show the theoretical profile, normalised and beam-convolved, assuming an A2 luminosity of 6 and 8 L$_\odot$ for PA=118 and 123°, respectively.
    The crosses are the normalised average of the intensity observed in the red- and blue- shifted emitting regions with its error bars, which are given by the difference of the red- and blue- shifted emitting regions (added to the 3$\sigma$ rms). 
    The dashed lines show the beam profile. 
    }
    \label{fig:OCS_line-intensity-profile}
\end{figure*}
%

\subsection{Theoretical OCS(19-18) line intensity} \label{subsec:model-line-intensity}

In order to compute the theoretical OCS(19-18) line intensity ($I_{OCS}$) profile, we assumed that the line is LTE populated (i.e., the excitation temperature is equal to the kinetic one, assumed to be equal to the dust temperature) and optically thin throughout the envelope, an assumption verified a posteriori, and for simplicity that the envelope is spherical. 
Under these assumptions:
\begin{equation}\label{eq:intensity}
    I_{OCS}(s) ~= \int_{LOS} {\rm x_{gas}(OCS)}(s,s') ~n_{gas}(s,s') ~h \nu ~A_{\nu} ~e^{-\frac{h \nu}{k_B~T_{gas}(s,s')}} ~{\rm ds'}
\end{equation}
where 
$I_{OCS}$(s) is the line intensity as a function of the projected distance from the envelope centre $s$, integrated along the line of sight ($s'$).
We used $I_{OCS}(s)$ normalised to 1, convolved with the synthesised beam, when comparing it with the observed profile, so it does not depend on the absolute values of $n_{gas}$ and ${\rm x_{gas}(OCS)}$, but only on their gradients.

The theoretical OCS(19-18) line intensity profile is shown in Fig. \ref{fig:OCS_abundance-intensity}, right panel. 
The sharp drop in the abundance with different luminosities is not converted into a similar drop in the theoretical intensity profile.
This is due to the combined effect of the decreasing column density and convolution at each point of the theoretical intensity with the telescope beam, which smears out the drop of ${\rm x_{gas}(OCS)}$ over a region of about 50 au.
To illustrate the beam smearing effect, Fig. \ref{fig:dif beam sizes} of Appendix A shows the theoretical intensity profile with a (hypothetical) telescope beam smaller or larger by a factor 10 with respect to that of the present observations (i.e. 0$\farcs$25). 
The radius at which the line intensity drops to 10\% of the peak moves from 90 to over 300 au with a 2 L$_\odot$ luminosity and from 210 to over 300 au with 15 L$_\odot$, for a beam size going from 0$\farcs$025 to 2.5$\farcs$, respectively.

\section{Discussion}\label{sec:discussion}

\subsection{Constraints on the bolometric luminosity of A2}\label{discuss-subsec:a2-lum}

In order to constrain the A2 luminosity, we used the reduced $\chi^2$ method, where $\chi^2$, in this case, is the sum of the squares of the difference between the predicted and measured normalized intensity profiles, divided by the square of the signal uncertainties.
The sum is then divided by the number of points (6) minus the freedom degrees (2).
We did not consider the innermost two data points, for two reasons:
(1) the dust absorption is expected to be high enough to lower the expected line intensity and (2) the temperature in the thick region of Eq. \ref{eq:temp-profile} may not be very reliable.
These issues are discussed in \S ~\ref{subsubsec:disc-lum-A1} and \ref{subsubsec:disc-temp-profile}, respectively, where it is shown that they do not impact the results of our analysis.


\subsubsection{Constraints on the density profile}\label{subsubsec:disc-dens-profile}
The theoretical OCS line intensity profile around A2 depends on the density profile $n \propto r^{-p}$.
To start with, we assumed to be the same as that found by \cite{Jorgensen2002} for the large-scale (24--24000 au) envelope.
In order to verify the validity of this assumption, we computed the $\chi^2$ for $p$ equal to 0, 1, 1.5, 1.8 \citep{Jorgensen2002} and 2.
Figure \ref{fig:dif-density-profiles} of Appendix A shows the results.
The lowest $\chi^2$, equal to 0.20, is obtained with $p$=1.5, namely that corresponding to a free-falling envelope.
The $\chi^2$ with $p$=1.8 is slightly larger, 0.65.
On the contrary, assuming $p$ equal to 0, 1 and 2 results in $\chi^2$ larger than 2.

In practice, the observed OCS line profile can constrain the density power law $p$, which turns out to be around the free-fall one ($p$=1.5).
We emphasize that the new value of $p$ is not in contrast with the \cite{Jorgensen2002} derived profile, which was obtained especially for the envelope at scale slightly larger than the ones probed by the FAUST observations.


\subsubsection{Our bona fide model}\label{subsubsec:bonafide-model}
Our bona fide model is the one obtained by using the density profile with $p$=1.5 and the temperature profile described in \S ~\ref{subsec:gas-profile}.
In the following subsections, we will explore how the adopted temperature profile impacts the derived A2 luminosity.

Taking the PA between 118 and 123°, as constrained in \S ~\ref{subsec:obs-results}, we obtained the theoretical, normalised and beam-convolved, OCS line intensity profile for different A2 luminosities and compared it with the measured one (reported in Fig. \ref{fig:pa-dif-pos}).
We then computed the reduced $\chi^2$ between the observations and theoretical predictions as a function of the A2 luminosity, for PA equal to 118 and 123°, respectively.
The results of the analysis are shown in Fig. \ref{fig:OCS_line-intensity-profile}.
The crosses in the lower panel of the figure represent the average of the OCS line intensity observed in the red- and blue- shifted emitting regions, while the error bars represent their difference (where the 3$\sigma$ rms on each point is also added).

The best $\chi^2$ is obtained for 6 L$_{\odot}$ for a PA of 118° and 8 L$_{\odot}$ for a PA of 123°.
We therefore conclude that the A2 luminosity is (7$\pm$1) L$_{\odot}$ when the bona fide model for the density and temperature profiles is used.

\subsubsection{Constraints on temperature profile}\label{subsubsec:disc-temp-profile}

In modelling the theoretical OCS line intensity profile we used the temperature profile derived by \cite{Jorgensen2002} for the entire envelope surrounding IRAS4A, which is based on observations that do not resolve its innermost region.
As described in \S ~\ref{subsec:gas-profile}, the temperature profile is made of two distinct regions, optically thick and optically thin, respectively.
While the absolute scaling with the 1/4 power of the luminosity in the thin region is rather safe, it is more problematic for the thick (i.e. innermost) region.
We, therefore, verified the impact of using a different luminosity dependence in the thick region, namely for $\rm 20 \leq r \leq 115$ au, running an additional model with a dependence on L$^{1/2}$ instead of L$^{1/4}$ (Eq. \ref{eq:temp-profile}).

The results are shown in Fig. \ref{fig:dif-lum-law} of Appendix A. 
Since the OCS gaseous abundance in these innermost regions is the same in the two models for luminosities $\geq$3 L$_\odot$, the line intensity is just slightly increased in the model with the L$^{1/2}$ dependence, because of the Boltzmann factor in the OCS line upper level population.
The effect is larger for luminosities $\leq$3 L$_\odot$, where the desorption occurs in the optically thick region. 
Since the luminosity of A2 is definitively larger than 3 L$_\odot$ (\S ~\ref{subsubsec:bonafide-model} and \ref{subsubsec:disc-dens-profile}), the results are not impacted by the actual temperature of the innermost region of the A2 envelope.
As a corollary, our choice to not consider the two innermost points of the A2 envelope is justified.

\subsection{Constraints on the bolometric luminosity of A1}\label{subsubsec:disc-lum-A1}

An analysis similar to that of A2 cannot be carried out for A1 because of the high continuum opacity of dust in the central region (\S ~\ref{subsubsec:ocs-a1}).
To constrain the luminosity of A1, we therefore just subtracted the A2 luminosity to the total A1+A2 one, (14.5$\pm$1.5) L$_{\odot}$ (see Sec. \ref{sec:source}).
Using the bona fide model, we obtain an A1 luminosity of (7.5$\pm$2.5) L$_{\odot}$.

\begin{figure}
    \centering
    \includegraphics[width=0.9\linewidth]{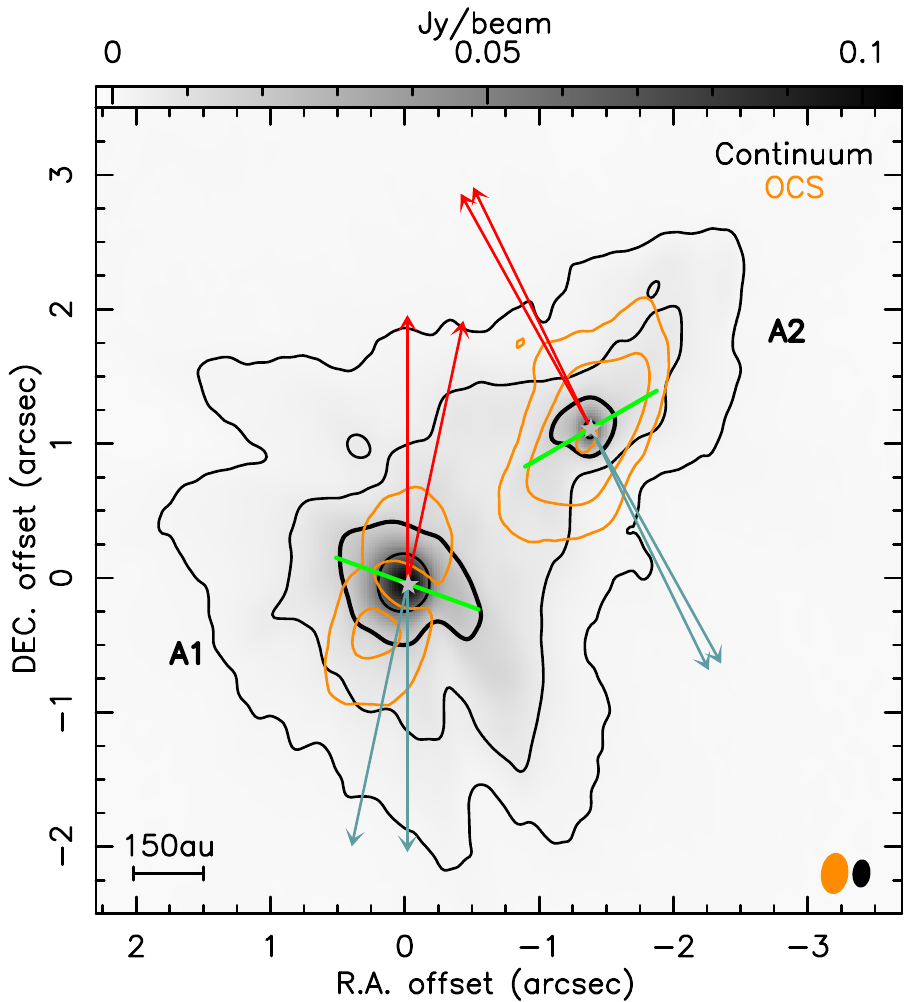}
    \caption{Moment 0 of OCS line (orange) and continuum (black) emission around A1 and A2. 
    OCS contours are at 5$\sigma$ ($\sigma$ = 3.5 mJy/beam km/s), 30$\sigma$ and 150$\sigma$. 
    Continuum contours are at 15$\sigma$ ($\sigma$ = 0.12 mJy/beam), 50$\sigma$, 150$\sigma$ (bold) and 500$\sigma$. 
    The green bar centred on A2 shows the size of the OCS emission $\geq$10\% of the maximum intensity, with a PA=120°. 
    In A1, the green bar has the same size but with a PA=70°, aligned with the A1 envelope. 
    Grey stars represent the position of A1 and A2 at the continuum peak emission. 
    Red and blue arrows represent red-shifted and blue-shifted outflow directions, respectively \citep{Chahine2024-cav}.
    The synthesised beams are represented by the orange and black ellipses in the bottom right corner of the map.}
    \label{fig:ocs-continuum}
\end{figure}

The very likely explanation of why OCS emission is observed towards A2 and not towards A1 is the high dust absorption around A1.
To test this hypothesis, in Fig. \ref{fig:ocs-continuum} we plotted the OCS overlapped with the continuum emission around the two sources. 
The green bar towards A2 represents where the OCS line intensity drops to 10\% of the maximum, in the envelope direction. 
The same green bar is then reported towards A1, in the direction of the envelope: it shows that it falls in a region where the continuum emission is larger than 150$\sigma$.
In A2, only the central beam falls into such large continuum emission contour where, by the way, our theoretical model overestimates the line intensity (Fig. \ref{fig:OCS_line-intensity-profile}, bottom panels), probably because of the dust absorption there.
In conclusion, the absence of OCS emission in the envelope of A1 is consistent with the hypothesis that it is completely absorbed by the dust.
As a corollary, our choice to not consider the two innermost points of the A2 envelope is justified.

\subsection{A1 and A2 luminosity: summary}

The comparison between the observed and predicted OCS line intensity profile provides a measurement of the A2 luminosity of (7$\pm$1) L$_\odot$.
Consequently, the A1 luminosity obtained subtracting the A2 luminosity to the total A1+A2 luminosity (14.5$\pm$1.5) L$_\odot$ results (7.5$\pm$2.5) L$_\odot$.

\begin{figure}
    \centering
    \includegraphics[width=0.9\linewidth]{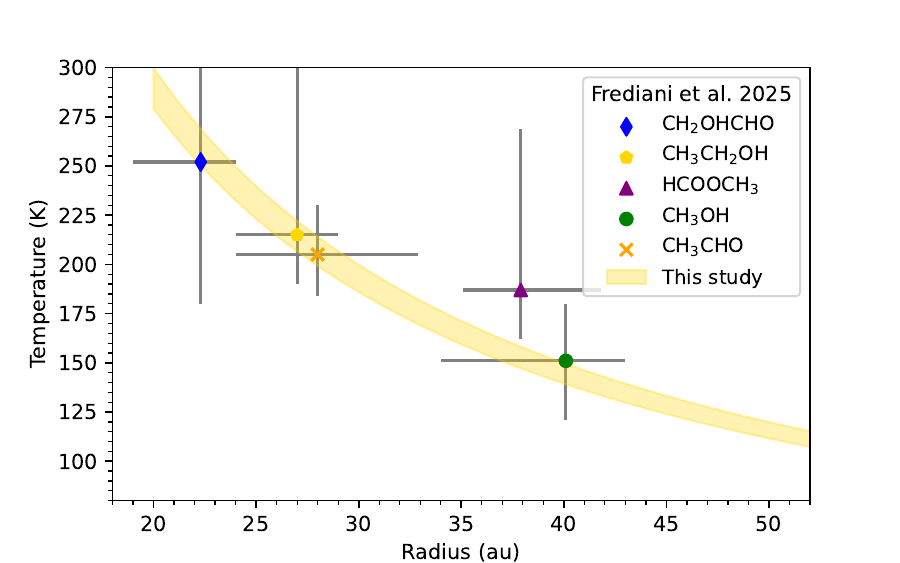}
    \caption{Theoretical temperature profile assuming (7$\pm$1) L$_{\odot}$ (yellow band) against the rotational temperatures of different molecules \citep[data points;][]{Frediani2025} around A2.}
    \label{fig:comp-frediani}
\end{figure}

To verify the reliability of the derived A2 luminosity, in Fig. \ref{fig:comp-frediani} we plot the theoretical temperature profile assuming (7$\pm$1) L$_{\odot}$ against the rotational temperatures of different molecules derived by \cite{Frediani2025} around A2.
The figure shows a good agreement between the temperatures predicted by our model and those derived by \cite{Frediani2025}, lending support to our derivation of the A2 luminosity.

The similarity of the A1 and A2 luminosity is in agreement with the study by \cite{Desimone2020-VLA} and \cite{Desimone2022-ices}, who showed that the sizes of the methanol emission in the centimetre, unaffected by dust absorption,  are similar in both sources. 
Like OCS, methanol is sublimated off the dust mantles, but in a smaller region than OCS because of the larger BEs \citep{Bariosco2025}.
Nonetheless, as in OCS, the luminosity regulates the sizes of the emission and the similarity of the ones derived from the CH$_3$OH line emission in A1 and A2 \citep{Desimone2020-VLA, Desimone2022-ices} militates for a similarity of the luminosity of the two sources.

\subsection{The importance of having the correct binding energy and prefactor distribution}\label{subsec:importance-BE}

In this work, we have shown that it is possible to measure the luminosity of the components of embedded protostars via high-resolution observations of the OCS line intensity profile.
The method is based on the underlying idea that OCS molecules are frozen at a certain distance from each binary component's centre because of their BEs.
Therefore, the BE is a crucial parameter of the method.
Actually, it is now clear that there is not a single value for the BE of a species but rather a BE distribution \citep[e.g.,][]{Tinacci2022-ammonia, Tinacci2023, bariosco_h2s, Bariosco2025, Groyne2025-BEDs}, as also shown for the OCS molecule in Sec. \ref{sec:OCS_BE-computations}.

Previous studies have reported the BE distribution of other molecules, notably that of methanol \citep{Bariosco2025} which can also be used to measure the luminosity following the method presented here, being CH$_3$OH a species formed prevalently on the grain surfaces \citep[e.g.,][]{Watanabe2002, Rimola2014}.  
However, because the methanol BEs are larger than those of OCS (e.g., the average methanol BE is 4255 K against 1215 K of OCS) the region where it is gaseous is smaller than that of OCS.
Therefore, it is more difficult to constrain the source luminosity using methanol, because it requires an higher spatial resolution. 
For example, in the A2 case with 7 L$_\odot$, the radius of the CH$_3$OH sublimation region would be $\leq$75 au against $\sim$190 au for OCS.
The necessity of choosing a molecule with a fairly low binding energy is an important parameter to use this method efficiently.

In dense gas, such as that close to the centre of an embedded protostar, molecules with low BEs sublimate farther away from the centre (where temperatures are lower) to freeze again onto the grain surfaces.
Statistically, therefore, the largest BE sites will end up being populated and, consequently, in the modelling what really matters is the BE of the populated sites.
Therefore, it is of paramount importance to have an accurate estimate of the BE distribution and not only a value, even less the average BE value.
To illustrate this point, we have estimated the luminosity of A2 assuming a BE and prefactor $\nu_{des}$ equal to the average value of the OCS distribution, computed in Sec. \ref{sec:OCS_BE-computations}, namely BE = 1215 K and $\nu_{des} = 8.9\times 10^{15}$ s$^{-1}$.
Using these values we found an A2 luminosity of less than 1 L$_\odot$, which would be totally unreasonable, because it would imply an A1 luminosity that is far too large (see, e.g., the discussion in \S ~\ref{subsubsec:disc-lum-A1}).

Finally it is also important to use the correct prefactor value (Eq. \ref{eq:k-des-ads}) associated with the relevant BE (Tab.\ref{tab:population_distrib}).
For example, when using the value often adopted in models by \cite{Wakelam2017BE}, namely BE = 2100 K and $\nu_{des} = 2\times 10^{12}$ s$^{-1}$, the resulting A2 luminosity would be 10 L$_\odot$, larger than that derived using our OCS BE distribution (while it should have been smaller, as the BE is larger than that used in our modelling).
The reason for this is the used prefactor, which is about four orders of magnitude too small with respect to the value obtained in the most recent computations.

Finally, we emphasize that the BE distribution dramatically impacts also the position of the molecules' snow-line in protoplanetary disks \citep[see, e.g., the case for water:][]{Tinacci2023}.
Specifically, instead of having a sharp increase/decrease of the molecule abundance in the disk, there will be a "diffuse" snow-line with a smooth transition solid/gaseous phase \citep[e.g.,][]{Boitard-Crepeau2025-terrestrialH2O}.
Likewise, the radius of the region where a molecule sublimates after a substantial luminosity burst of a protostar, such as in B335 where a burst of 5-7 times the luminosity has been observed \citep{Bjerkeli2023-B335burst, Lee2025-B335}, depends on the molecule BE distribution and the interpretation of the observations depends on the assumed value.
In this case, one has also to consider that the sublimated molecules will re-condense in time, so that the gaseous abundance depends on the timescale of freeze-out as well as the saturation of the BE bins (see description in \S ~\ref{subsec:abu-profile}) and one has to explicitly solve the molecular abundance equations as a function of time.

\section{Conclusions}\label{conclusions}


We presented a new method for estimating the luminosity of embedded protostars in binary or, more generally, multiple systems which are unresolved in the IR, where the classical method of measuring it from the SED cannot be applied.
The new method is based on the fact that gaseous carbonyl sulphide is largely due to the thermal sublimation of grain mantles, so that comparing the observed and theoretical OCS line intensity profile in the envelope heated by the central embedded source provides an estimate of the protostar luminosity.
To make the estimate reliable, a good knowledge of the OCS binding energy (BE) distribution is needed; likewise,  the observations need to correctly sample the line intensity profile originating in the envelope.

Since the OCS BE distribution was not known, we carried out new quantum mechanics (QM) calculations, following up a method developed by our group in previous works \citep{Tinacci2022-ammonia, Tinacci2023, bariosco_h2s, Bariosco2025}.
The OCS BE distribution is well described by a gaussian whose mean BE value is 10.1 kJ/mol (1215 K) and standard deviation is 4 kJ/mol (481 K).
The computed BEs range from 1.9 kJ/mol (228 K) to 20.3 kJ/mol (2436 K).

We applied the new proposed method to the protobinary system IRAS4A, whose two components are A1 and A2.
We reported new high spatial resolution ($\sim$50 au) observations of the OCS(19-18) line at 231 GHz, obtained in the framework of the Large Program ALMA FAUST \citep{Codella2021}.
While the OCS emission around A2 extends to a radius of about 200 au,  it is completely absorbed towards A1 by the dust that surrounds it, whose emission is at least three times larger in spatial extent than in A2.
We therefore applied the method to A2 after disentangling the OCS emission of the A2 envelope from that of the cavities of the outflows emanating from A2, based on the work by \citet{Chahine2024-cav}.
The observed OCS line intensity profile of the A2 envelope was then compared to the theoretical one, which greatly depends on the A2 luminosity.

Via a $\chi^2$ analysis we found that the A2 luminosity is (7$\pm$1) L$_\odot$.
Subtracting it from the previously measured total luminosity of the IRAS4A binary system, (14.5$\pm$1.5) L$_\odot$, we also constrained the A1 luminosity, which results in (7.5$\pm$2.5) L$_\odot$.
The derived luminosities are in agreement with previous dust-absorption free observations of the methanol in the A1 and A2 hot corinos, whose sizes are similar \citep{Desimone2020-VLA}, and the gas temperature measured in the 20--40 au region around A2 \citep{Frediani2025}.

We discussed the dependence of our results on the adopted theoretical temperature and density profiles of the A2 envelope.
We found that the density profile is constrained by the OCS observations and it is in agreement with that of a collapsing envelope.
Furthermore, the uncertainty in the temperature profile in the innermost envelope (radius $\leq$ 70 au) of A2 does not affect the results.

Finally, we discussed the importance of having a reliable OCS BE and prefactor distributions in obtaining reliable luminosities. 

The method presented in this work can also be applied to and tested in single sources where an independent estimate of the luminosity exists.
Unfortunately, we do not have such a source at hand but we plan to obtain new OCS observations to this end. 
In principle, our method can also be applied with other molecules for which the BE distribution is known.
However, OCS is an optimal molecule because it is mostly formed on the grain surfaces and has a relatively low BE, which implies that the sublimation region is relatively large and can be resolved by interferometric millimetre observations, as those presented here.
That is not the case, for example, of methanol, whose larger BEs \citep{Bariosco2025} make the emitting region smaller and, hence, more difficult to be spatially resolved.

We emphasize that the better the spatial resolution of the observations the more reliable the estimate of the luminosity of embedded protostars in binary or multiple systems.
Also, ideally, observations in the centimetre wavelengths, where dust absorption is minimal, hold the promise to provide the most reliable estimate of the luminosity of the two components of a binary (or multiple) system where one of the two has a sufficiently large dust mass to completely obscure molecular line emission at millimetre wavelengths.
This is the case of several young systems, such as NGC1333 IRAS4A, studied here, and IRAS16293-2422 \citep{Bottinelli2004-i16293PdB, Jorgensen2018}.

We conclude stressing the paramount importance of coupling astronomical observations with QM chemistry studies to fully exploit the message that we receive from the observed data.

\section*{Acknowledgments}
This article makes use of the following ALMA data: ADS/JAO.ALMA2018.1.01205.L (PI: S. Yamamoto). ALMA is a partnership of the ESO (representing its member states), the NSF (USA) and NINS (Japan), together with the NRC (Canada) and the NSC and ASIAA (Taiwan), in cooperation with the Republic of Chile. The Joint ALMA Observatory is operated by the ESO, the AUI/NRAO, and the NAOJ.
This project has received funding from the European Research Council (ERC) under the European Union's Horizon 2020 research and innovation program, for the Project “The Dawn of Organic Chemistry” (DOC), grant agreement No 741002, and from the Marie Sklodowska-Curie for the project ”Astro-Chemical Origins” (ACO), grant agreement No 811312.
P.U. acknowledges the support from Project CH4.0 under the MUR programme ‘Dipartimenti di Eccellenza 20232027’ (CUP: D13C22003520001).
E.B. acknowledges contribution of the Next Generation EU funds within the National Recovery and Resilience Plan (PNRR), Mission 4 - Education and Research, Component 2 - From Research to Business (M4C2), Investment Line 3.1 - Strengthening and creation of Research Infrastructures, Project IR0000034 – “STILES - Strengthening the Italian Leadership in ELT and SKA”, and 
the support of the Italian Ministry for Universities and Research under the Italian Science Fund (FIS 2 Call - Ministerial Decree No. 1236 of 1 August 2023).
Cl.Co., L.P. and Gi.Sa. acknowledge the PRIN-MUR 2020  BEYOND-2p (Astrochemistry beyond the second period elements, Prot. 2020AFB3FX), 
the project ASI-Astrobiologia 2023 MIGLIORA (Modeling Chemical Complexity, F83C23000800005), and 
the INAF-GO 2023 fundings PROTO-SKA (Exploiting ALMA data to study planet forming disks: preparing the advent of SKA, C13C23000770005). 
Cl.Co. and L.P. acknowledge financial support under the National Recovery and Resilience Plan (NRRP), Mission 4, Component 2, Investment 1.1, Call for tender No. 104 published on 2.2.2022 by the Italian Ministry of University and Research (MUR), funded by the European Union – NextGenerationEU-Project Title 2022JC2Y93 Chemical Origins: linking the fossil composition of the Solar System with the chemistry of protoplanetary disks – CUP J53D23001600006 – Grant Assignment Decree No.962 adopted on 30.06.2023 by the Italian Ministry of Ministry of University and Research (MUR). 
L.P. acknowledges the INAF Grant 2022 “Chemical Origins” and the INAF Grant 2024 "ICES". 
Gi.Sa. acknowledges the support from the INAF-Minigrant 2023 TRIESTE ("TRacing the chemIcal hEritage of our originS: from proTostars to planEts”).
M.B. acknowledges the support from the European Research Council (ERC) Advanced Grant MOPPEX 833460.
L.L. acknowledges the support of DGAPA-UNAM PAPIIT grants IN108324 and IN112820, CONAHCyT grant CF-263356 and SECIHTI grant CBF-2025-I-109.
Finally, we warmly thank Patrice Theulé and Timea Csengeri for very useful discussions and the manuscript's reviewer, Neil Evans for comments that helped to improve article.
Part of the data reduction/combination presented in this paper was performed using the GRICAD infrastructure (https://gricad.univ-grenoble-alpes.fr).

\section*{Data Availability}

The raw data are available on the ALMA archive at the end of the proprietary period (ADS/JAO.ALMA\#2018.1.01205.L).



\bibliographystyle{mnras}
\bibliography{SauryGuillaume} 

\newpage
\appendix
\section{Additional figures}
In this appendix we report additional figures, presented and discussed in the main body.

\begin{figure*}
    \centering
    \begin{tabular}{c}
         \includegraphics[width=0.6\linewidth]{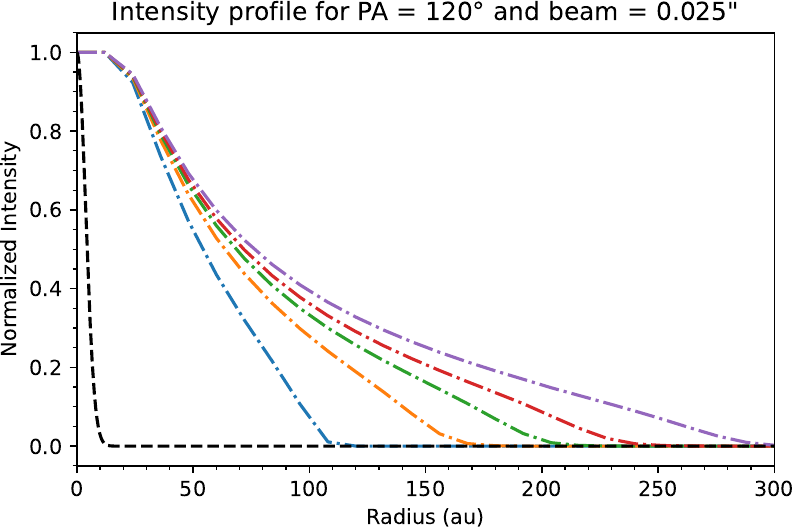} \\ 
         \includegraphics[width=0.6\linewidth]{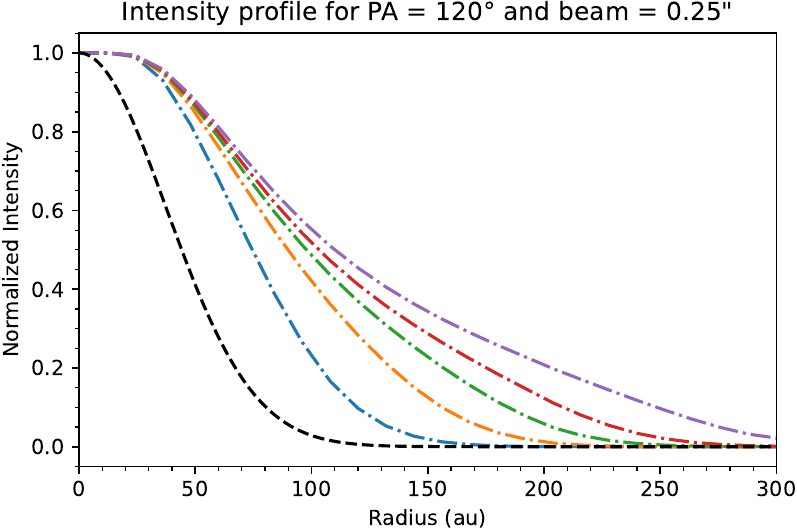} \\ 
         \includegraphics[width=0.6\linewidth]{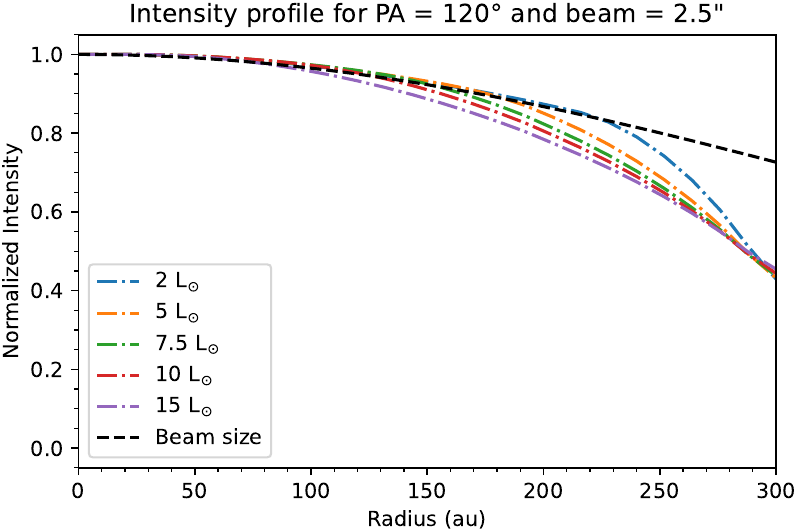} 
    \end{tabular}
    \caption{Theoretical line intensity profiles at a PA of 120° derived using beam sizes:
    0$\farcs$025, 0$\farcs$25 and 2.5$''$, as marked in the panels.}
    \label{fig:dif beam sizes}
\end{figure*}

\begin{figure*}
    \centering
    \begin{tabular}{cc}
         \includegraphics[width=0.37\linewidth]{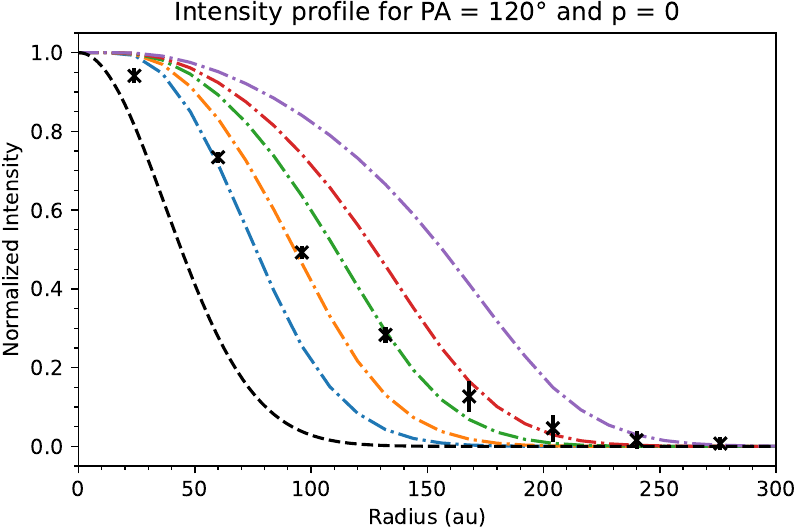} 
         \includegraphics[width=0.37\linewidth]{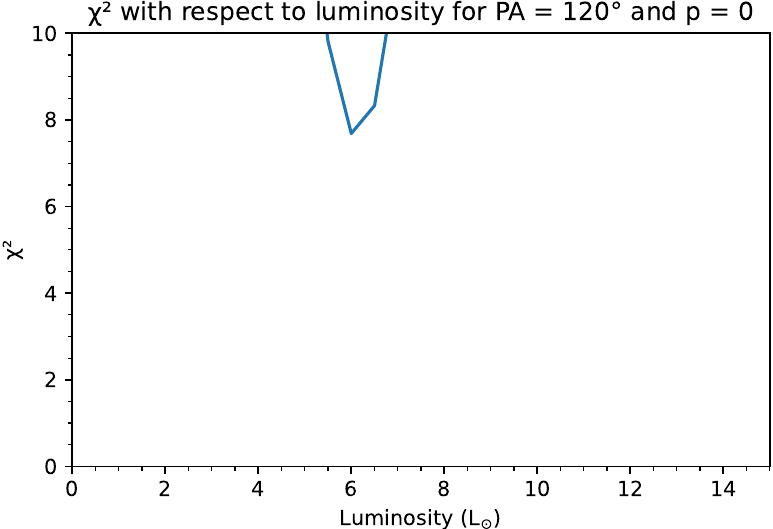} \\ 
         \includegraphics[width=0.37\linewidth]{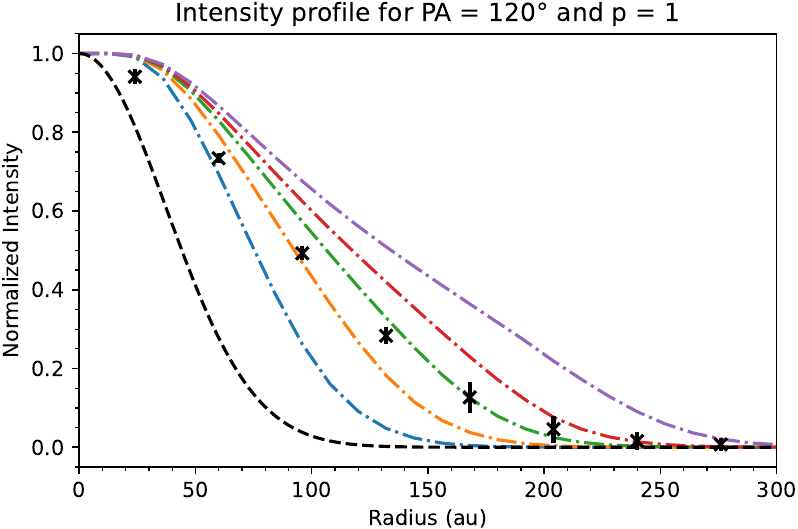} 
         \includegraphics[width=0.37\linewidth]{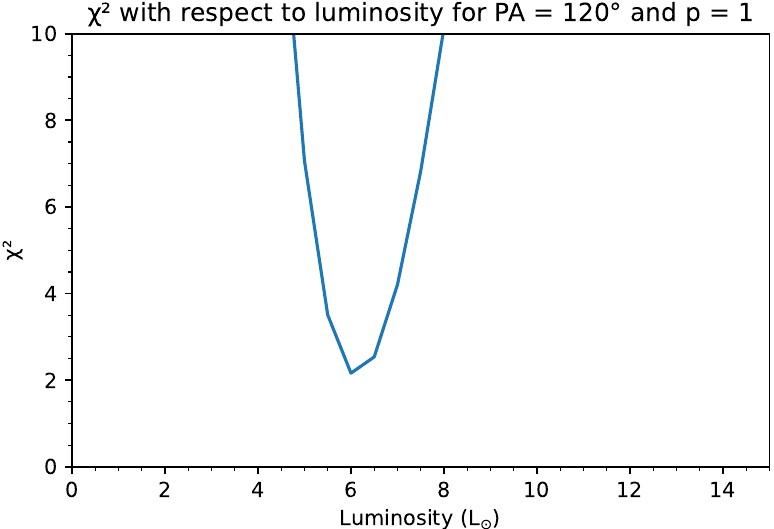} \\ 
         \includegraphics[width=0.37\linewidth]{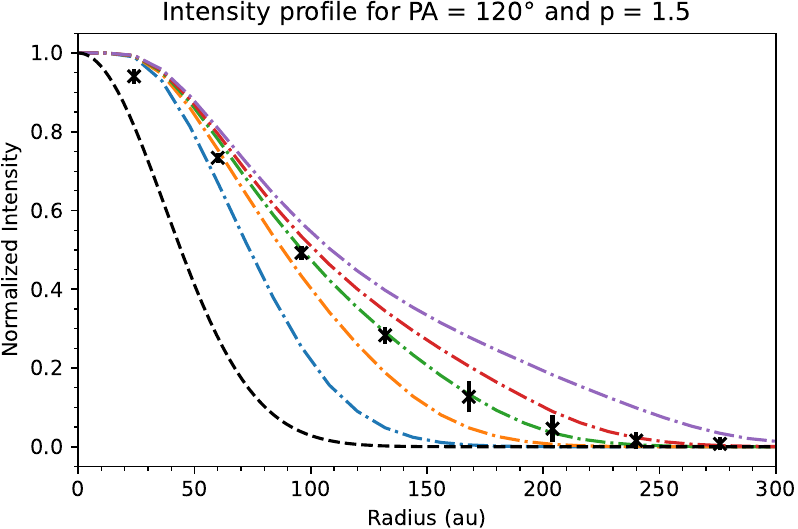} 
         \includegraphics[width=0.37\linewidth]{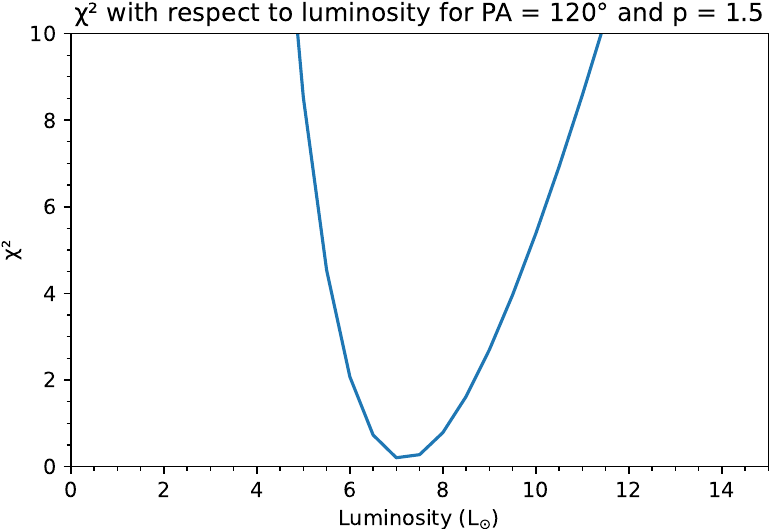} \\ 
         \includegraphics[width=0.37\linewidth]{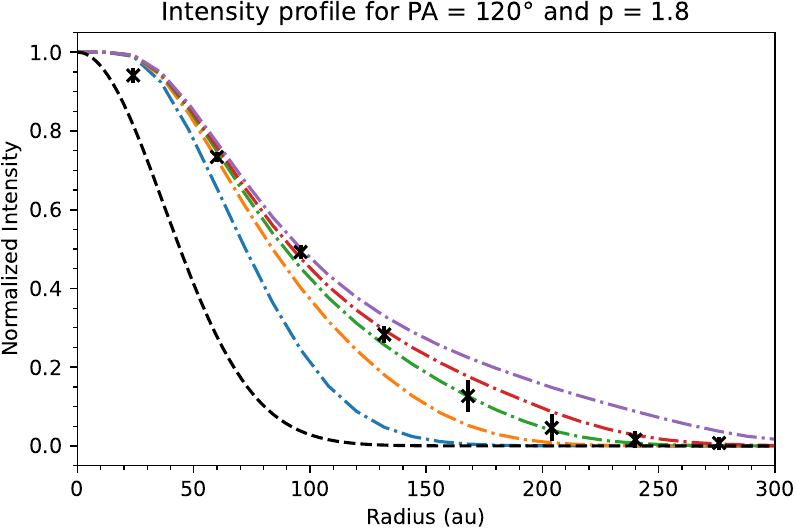} 
         \includegraphics[width=0.37\linewidth]{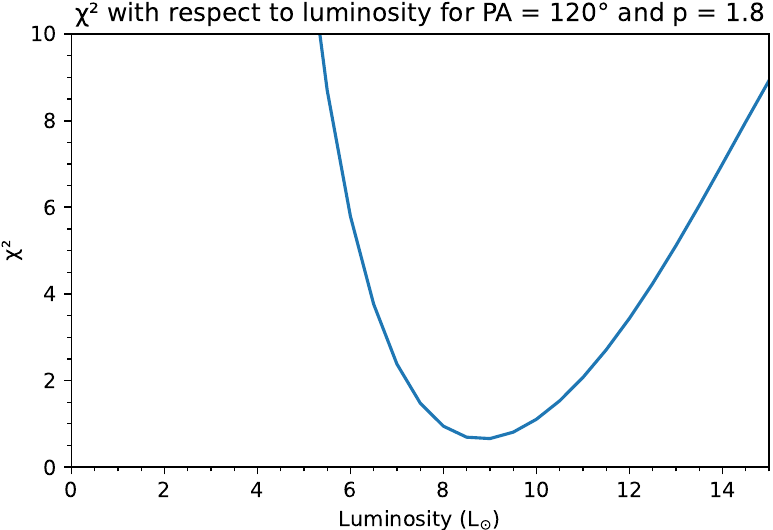} \\ 
         \includegraphics[width=0.37\linewidth]{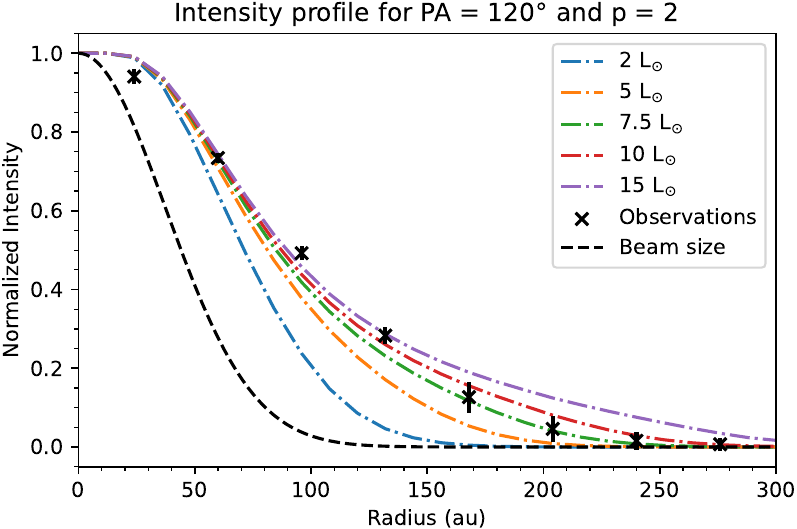} 
         \includegraphics[width=0.37\linewidth]{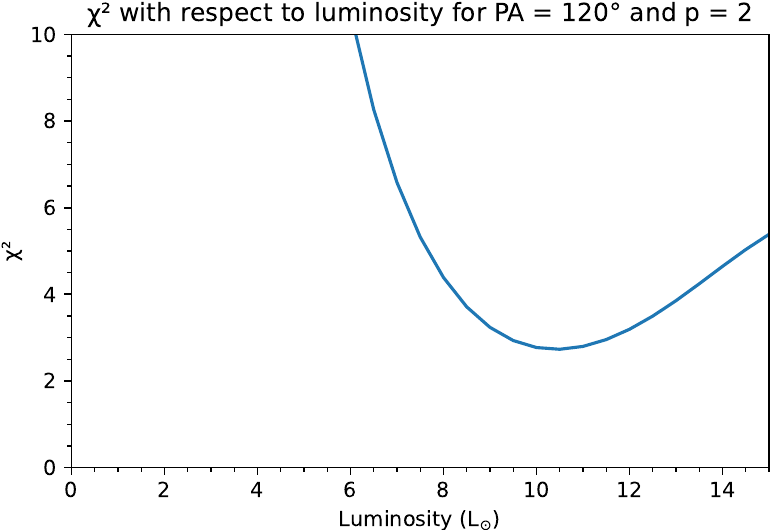} \\ 
    \end{tabular}
    \caption{\textit{Left panels:} OCS line intensity profiles with different density profiles. 
    The density profiles evolves as $n \propto r^{-p}$ with p = 0, 1, 1.5, 1.8 and 2.
    \textit{Right panels:} ${\chi}^2$ as a function of the A2 luminosity, for PA=120° (see \S ~\ref{subsubsec:ocs-a2}) without taking account the two first points of the observations for each intensity profile.}
    \label{fig:dif-density-profiles}
\end{figure*}

\begin{figure*}
    \centering
    \begin{tabular}{cc}
        \includegraphics[width=0.48\linewidth]{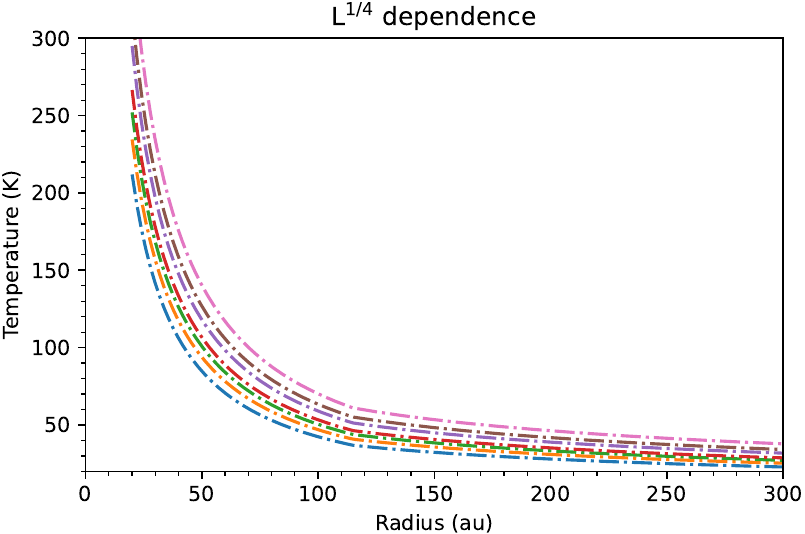} & \includegraphics[width=0.48\linewidth]{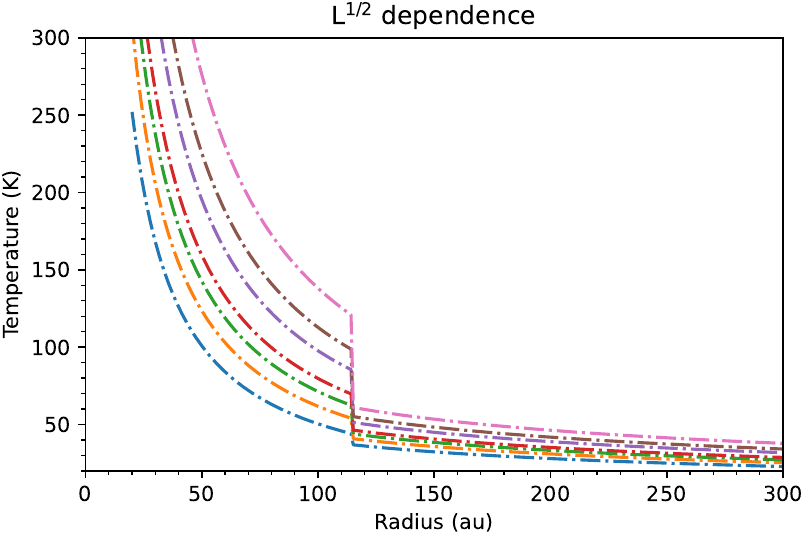} \\
        \includegraphics[width=0.48\linewidth]{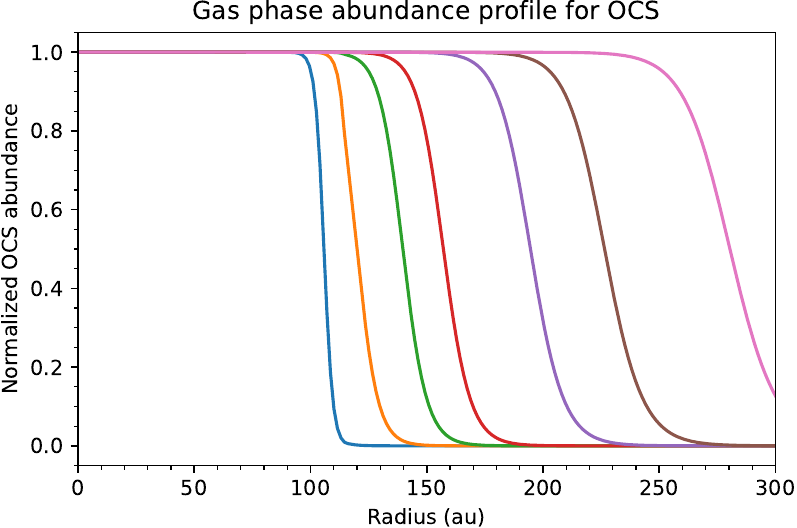} & \includegraphics[width=0.48\linewidth]{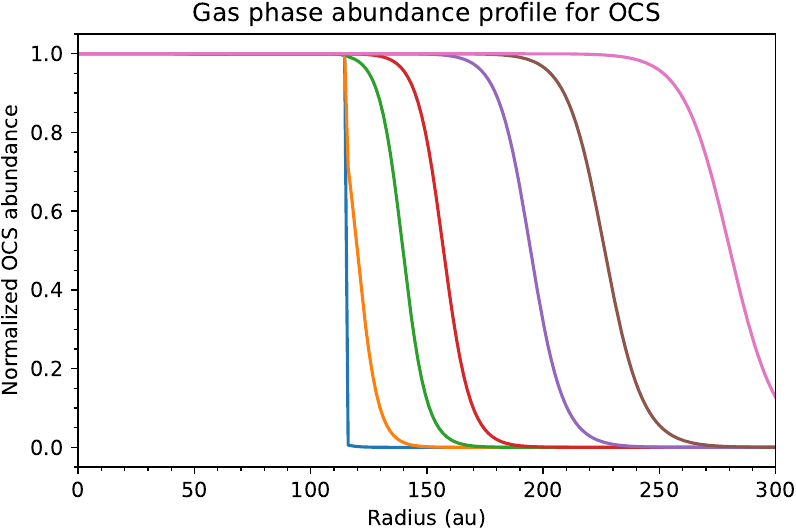} \\
        \includegraphics[width=0.48\linewidth]{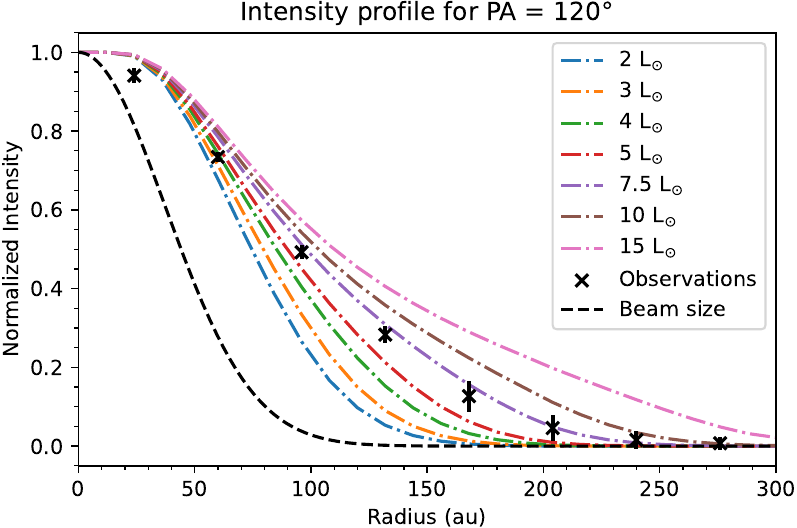} & \includegraphics[width=0.48\linewidth]{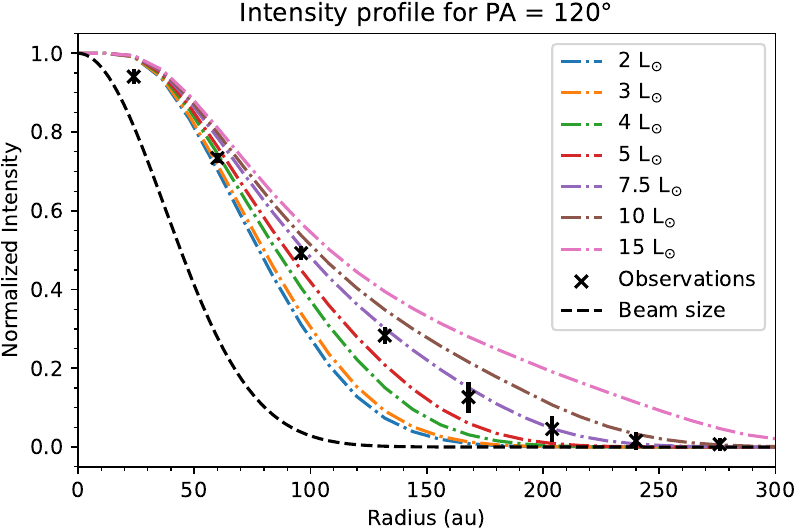} 
    \end{tabular}
    \caption{Temperature, abundance and intensity profiles for different power laws of the luminosity in the optically thick region. 
    Left and right columns are obtained with dependence on L$^{1/4}$ and L$^{1/2}$, respectively.}
    \label{fig:dif-lum-law}
\end{figure*}

\bsp	
\label{lastpage}
\end{document}